\documentclass{aa}  

\usepackage{graphicx}
\usepackage{txfonts}
\usepackage{hyperref}
\usepackage{dblfloatfix}

\newcommand{\msol}{M$_\odot$}
\newcommand{\rsol}{R$_\odot$}
\newcommand{\lsol}{L$_\odot$}
\newcommand{\teff}{$T_{\textrm{eff}}$}
\newcommand{\logg}{$\log\,g$}
\newcommand{\he}{$\log(N_{He}/N_{H})$}

\newcommand{\kms}{$\mathrm{km\,s^{-1}}$}

\begin{document}

   \title{Spectroscopic follow-up of hot subdwarf variables found in ZTF}

   \subtitle{Atmospheric and fundamental properties of radial-mode sdB pulsators}

   \author{Corey W. Bradshaw
          \inst{1}\fnmsep\thanks{Corresponding author; \texttt{corey.bradshaw@uni-hamburg.de}}
          \and
          Thomas Kupfer\inst{1,2}
          \and
          Alekzander R. Kosakowski\inst{2,3}
          \and
          Brad N. Barlow\inst{3}
          \and
          Matti Dorsch\inst{4}
          }

   \institute{Hamburger Sternwarte, University of Hamburg, Gojenbergsweg 112, 21029 Hamburg
         \and
             Department of Physics and Astronomy, Texas Tech University, Lubbock, TX 79409, USA
         \and
             Department of Physics and Astronomy, University of North Carolina at Chapel Hill, Chapel Hill, NC 27599, USA
         \and
            Institut f{\"u}r Physik und Astronomie, Universit{\"a}t Potsdam, Haus 28, Karl-Liebknecht-Str. 24/25, 14476 Potsdam, Germany
            }

   \date{Received Month dd, yyyy; accepted Month dd, yyyy}

 
  \abstract
   {Hot subdwarf variables (sdBVs) that display large-amplitude ($>$1\%), short-period variability, as a result of radial-mode pulsations, have recently become objects of interest as they show unique properties among the sdBV classes. Since the discovery of objects such as Balloon 090100001 and CS\,1246, twelve more have been discovered in the Zwicky Transient Facility (ZTF) survey that display similar characteristics. However, due to lack of broad spectroscopic investigations, it remains unclear whether these objects constitute a distinct class of radial-mode dominant sdBVs that share common atmospheric and fundamental properties.}
   {Here we aim to spectroscopically define these peculiar sdBVs as a population and determine if they constitute a unique class of pulsating hot subdwarfs or share similar properties with preexisting classes.}
   {We collected low-resolution spectroscopy on a sample of sdBVs discovered in the ZTF survey, including time-series observations. We fitted the spectra to a grid of theoretical models to determine their mean effective temperature, surface gravity and helium abundance and any corresponding variability. We then use these properties to estimate the mass, radius and luminosity using a spectral energy distribution fitting method.}
   {We show that the resulting properties are similar to the radial-mode dominant sdBVs, Balloon 090100001 and CS\,1246, and that they are distinguishable from other similar radial-mode pulsators, such as blue large-amplitude pulsators. We find that these stars, on average, have mean effective temperatures of 28\,300~K and surface gravity measurements of {\logg}=5.56, with changes in these parameters on the order of 1000~K and 0.10 dex, respectively. The location of these stars on the {\teff} -- {\logg} plane places them on the boundary region between the low-amplitude, multi-periodic V361 Hya and V1093 Her stars, where the hybrid DW Lyn pulsators lie. The masses and radii of the majority of the sdBVs in our sample align with canonical-mass sdB properties. 
   } 
   {}


   \maketitle
%

\section{Introduction}
\label{sec:intro}

Hot subluminous stars, or hot subdwarfs, are a unique class of evolved stellar objects which have spectral types O or B (sdOs {\&} sdBs) and luminosities which lie below the main sequence and above the white dwarf (WD) track, at the blue end of the horizontal branch, known as the extreme horizontal branch (EHB) \citep{Heber2009, Heber2016, Heber_2024}. Most hot subdwarfs are low-mass ($\approx$0.50~{\msol}) helium core-burning stars, with thin hydrogen envelopes \citep{Heber2016}. Particularly, sdB stars have effective temperatures ({\teff}) in the range of $20\,000$ to $30\,000$~K and typical surface gravities between {\logg} = 5 and 6 (cgs) \citep{Heber1986}. These objects are thought to form through interaction with a binary companion via common envelope evolution (CEE) or stable Roche lobe overflow (RLOF) \citep{Han2002,Han2003}. In fact, a significant fraction of hot subdwarfs are observed in binary systems \citep{Maxted2001,stark_2003}.

Pulsations have been both theorized and detected in hot subdwarfs, leading to the classification of several types of sdB variables (sdBVs). These pulsations were initially attributed to non-radial stellar oscillations driven by the $\kappa$-mechanism associated with an opacity bump produced by the partial ionization of iron-group elements \citep{Charp1996,Charp1997}. Shortly after, low-amplitude, multi-periodic variability was observed in sdBs, in what were called V361 Hya and V1093 Her stars \citep{kil_1997,Green_2003}. The rapidly pulsating V361 Hya (sdBV$_{r}$) stars have {\teff} $\gtrsim$~$28\,000$~K and show pressure mode ($p$-mode) oscillations on the timescale of a few minutes \citep{kil_1997}. The slowly pulsating V1093 Her (sdBV$_{s}$) stars were reported with {\teff} $\lesssim$~$28\,000$~K and show longer period gravity mode ($g$-mode) pulsations on the timescale of $45$~min - $2$~hrs \citep{Green_2003}. These classes of sdBVs typically show small amplitudes of only up to a few millimagnitude (mmag). However, following these discoveries, sdBVs have been found that have large-amplitude pulsations of a few percent or greater. 

With the discovery of Balloon 090100001 (BA09), these pulsating sdBs were shown to also experience not only large-amplitude pulsations, but possess the capability of producing both rapid $p$-mode and slower $g$-mode oscillations \citep{Oreiro_2004,Baran_2005}. This star was shown to have at least 50 pulsation modes across both the theoretical sdB $p$ and $g$-mode regimes and spectroscopically resides on the boundary region between the sdBV$_{r}$ and sdBV$_{s}$ stars, creating the hybrid DW Lyn (sdBV$_{rs}$) class \citep{Baran_2006,schuh_2006}. A concise overview of the proposed nomenclature for sdBVs is presented in \citet{sdBV_names_2010}. The dominant, large-amplitude pulsation in BA09 was attributed to a radial-mode oscillation from time-series spectroscopy \citep{Telting_2006, Telting_2008}, and fully constrained using multi-band photometry \citep{Baran_2008,Charp_2008}. Recently, another hybrid sdBV$_{rs}$, BPM 36430, has been discovered with a dominant radial mode, and has some similarities to BA09 \citep{Krzen_2022,Smith_2022,Kirby_2023}.

\citet{Barlow_2010} discovered the first short-period sdBV whose frequency spectrum was dominated by a large, single-mode oscillation that was reported to have a semi-amplitude of $\approx30$~mmag and a period of $371.707\pm0.002$~s named CS\,1246. Through time-series spectroscopy this star was determined to be radially pulsating by observing sinusoidal radial velocity (RV), {\teff} and {\logg} variations, consistent with uniform expansion and contraction of the stellar surface. Atmospheric measurements for this object resulted in an average {\teff} of $28\,450\pm700$~K with an average {\logg} of $5.46\pm0.11$. Using the Baade-Wesselink method, CS\,1246 was found to have a mass of $0.39_{-0.13}^{+0.30}$~M$_{\odot}$ and a radius of $0.19 \pm 0.08$\,$R_\odot$ and constitutes the first single-mode, radially pulsating sdBV, with a short pulsation period and large amplitude of a few tens of mmag. 

Since the discovery of CS\,1246, 12 objects have been discovered as part of the Zwicky Transient Facility (ZTF) high-cadence Galactic plane survey, which have similar photometric properties as CS\,1246, BA09 and BPM 36430 \citep{ZTF_sdbv_2021}. These stars were reported to show only one dominant oscillation mode that was assumed to be radial, based on the period, amplitude and sinusoidal nature of the light curves, all of which heavily resemble CS\,1246. These $12$ objects have pulsation amplitudes of $40$ - $145$~mmag and pulsation periods of $5.8$ - $16$~min, with the majority of them having periods of $\approx6$~min, aligning very closely to that of CS\,1246. Very little is known about the atmospheric and evolutionary properties of these radial-mode dominant sdBVs (rm-sdBVs) as a class, since few spectroscopic observations have been carried out and no theoretical stellar evolutionary studies have been performed. 

Additionally, large-amplitude, radial-mode pulsations have been discovered in other stars which have observational characteristics of sdBs. These include both blue large-amplitude pulsators (BLAPs) and high-gravity BLAPs. \citet{OW_2017} as part of the Omega White (OW) survey reported observations of a hot, large-amplitude, short-period pulsator with unusual characteristics, which they dubbed a peculiar $\delta$-Scuti type star. Shortly thereafter, an initial sample of 14 stars, which included this object, were formally classified by the Optical Gravitational Lensing Experiment (OGLE) survey as BLAPs, and were reported to have pulsation periods in the range of 20 - 40 min, with amplitudes between 200 - 400 mmag \citep{BLAPs_2017}. Preliminary spectroscopic results for four objects in this sample showed {\teff} and {\logg} values of $\approx26\,000 - 30\,000$~K and $\approx4.50$, respectively. The number of BLAPs continues to grow today with currently over 100 candidate objects reported, about two dozen of which being confirmed spectroscopically. Consequently, the known pulsation period range has grown to somewhere between 3 and 60 min \citep{BLAPs_2023,BLAPs_2024,Obs_BLAPs_2024}. 

The high-gravity BLAPs were first discovered in ZTF and were observed to have shorter pulsation periods in the range of $\approx 3 - 8$~min and amplitudes of $\approx100 - 300$~mmag, with {\teff} and {\logg} values of $31\,000 - 34\,000$~K and $\approx5.50$, respectively \citep{Kupfer_2019}. \citet{Obs_BLAPs_2024} has recently shown that the originally discovered BLAPs and the high-gravity BLAPs form a homogeneous class of stars, and we therefore drop the distinction between the two groups throughout this work. Based on these results, it appears that the rm-sdBVs share similar properties with BLAPs, raising questions in the literature of whether the two classes are related (see \citealt{ZGP_BLAP_2022,OW_BLAPs_2022,koen_2024,Wu_2025}). 

This paper presents photometric and spectroscopic results for all objects in the sample of ZTF-sdBVs, in order to classify each and characterize their pulsational, atmospheric, and fundamental properties. Recently released ZTF photometry for the original sample of 12 ZTF-sdBVs is used to derive precise periods and multi-color amplitudes, as well as search for any additional pulsation frequencies. We also report five additional rm-sdBV candidates found in ZTF that display similar photometric characteristics as the original sample. Spectroscopic observations were carried out on all 12 stars in the original sample, plus one from the set of newly discovered objects, using the SOAR/Goodman, Lick/Shane-Kast and Gemini/GMOS-S instruments to measure the mean atmospheric properties ({\teff}, {\logg}, and {\he}) of each. Fully phase-resolved spectroscopic measurements were performed on several of these objects to observe any variability of their atmospheric properties and characterize their pulsations. We then make use of these results to estimate the fundamental properties (mass, radius and luminosity) for the observed sample. In Section~\ref{sec:data} we describe the complete set of photometric and spectroscopic data used in this study and in Section~\ref{sec:analysis} we detail the methods of our analysis. Section~\ref{sec:results} reports the results of this analysis in order to classify these objects and begin the characterization of these peculiar sdBVs. We discuss these results in Section~\ref{sec:disc} and provide closing remarks and statements on future studies for these objects in Section~\ref{sec:conc}.

\section{Data}
\label{sec:data}

\begin{table}
\centering
\caption{Photometric and spectroscopic observations.}
\label{tab:data}
\begin{tabular}{lcc|cc}
\hline
\text{Object} 
& \multicolumn{2}{c|}{Photometry} 
& \multicolumn{2}{c}{Spectroscopy} \\
& $N_{\rm obs}(g)$ & $N_{\rm obs}(r)$ 
& Instrument & $N_{\rm spectra}$ \\
\hline
ZTF-sdBV1 & 375 & 974 & Goodman* & 63 \\
ZTF-sdBV2 & 276 & 874 & Goodman & 3 \\
ZTF-sdBV3 & 843 & 1123 & Kast & 3 \\
ZTF-sdBV4 & 180 & 463 & Goodman* & 73 \\
ZTF-sdBV5 & 623 & 903 & Kast & 3 \\
ZTF-sdBV6 & 157 & 610 & GMOS* & 139 \\
ZTF-sdBV7 & 573 & 1527 & Goodman* & 260 \\
ZTF-sdBV8 & 129 & 541 & GMOS* & 70 \\
ZTF-sdBV9 & 227 & 703 & Goodman & 3 \\
ZTF-sdBV10 & 433 & 990 & Kast & 2 \\
ZTF-sdBV11 & 300 & 820 & Goodman & 2 \\
ZTF-sdBV12 & 190 & 688 & Goodman* & 77 \\
ZTF-sdBV13 & 284 & 697 & -- & -- \\
ZTF-sdBV14 & 634 & 1034 & -- & -- \\
ZTF-sdBV15 & 1165 & 1242 & -- & -- \\
ZTF-sdBV16 & 312 & 363 & Goodman* & 273 \\
ZTF-sdBV17 & 658 & 1065 & -- & -- \\
\hline
\end{tabular}
\tablefoot{Total number of both photometric and spectroscopic observations that were used in the analysis. For spectroscopic data, the utilized instrument is listed, and targets observed in time-series are denoted by an asterisk (*).}
\end{table}

\subsection{Photometry}
\label{sec:data:phot}

Photometric data for all stars presented here were collected from the Zwicky Transient Facility \citep{ZTF_2020}. ZTF is a large-area sky survey which utilizes the Palomar 48-inch (P48) telescope. This data covers approximately a 6~yr baseline in both the $g$ and $r$ bands using 30~s exposures. The resulting number of observations used in this study for each band are listed in Table~\ref{tab:data}. This data allows for the determination of a precise pulsation period, as well as pulsation amplitudes in two bands for the dominant oscillation mode of each target. 
The \citet{ZTF_sdbv_2021} sample of 12 ZTF-sdBVs were discovered in ZTF DR3 and DR4 and have continued to be observed through recent data releases. We identified an additional five objects (ZTF-sdBV13 - ZTF-sdBV17) as part of a search for short period binaries within the \textit{Gaia} EDR3 white dwarf catalog \citep{gentilefusillo2021} using data from ZTF DR10. A detailed description of the image processing and data reduction for ZTF can be found in \citet{ztf_reduc_2019}. The photometric data was retrieved from the ZTF database using the \texttt{lightcurve} module in the python package \texttt{ztfquery} \citep{ztf_query_2018}. 

\begin{figure}
    \includegraphics[width=\linewidth]{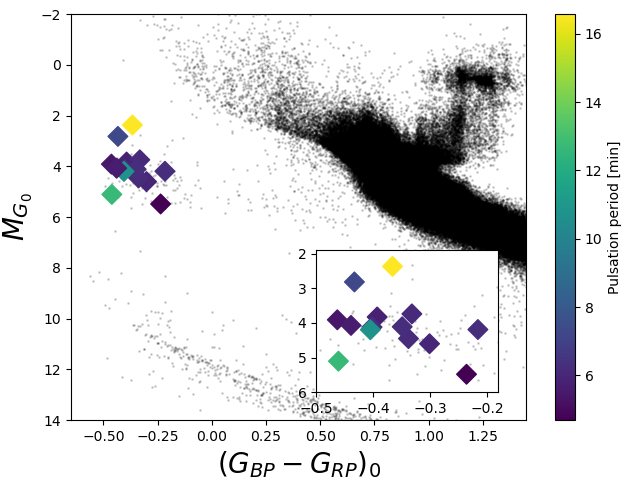}
    \caption{Color magnitude diagram constructed for objects in the ZTF-sdBV sample which have reasonable parallax measurements, using \textit{Gaia} photometric $G$, $G_{BP}$ and $G_{RP}$ color-corrected magnitudes. Each object is color-mapped with its corresponding pulsation period, which is scaled to the observed range of pulsation periods for the entire sample.}
    \label{fig:CMD}
\end{figure}

We collect parallax measurements ($\varpi$) for each star from \textit{Gaia} DR3 (\citealt{Gaia}, \citealt{Gaia_DR3}) as well as $G_{BP}-G_{RP}$ color indices and $G$ band magnitudes in order to place our sample on a color-magnitude diagram (CMD). The background star sample was queried using the parameters presented in \citet{Gaia_HRD_2018} such that a clean selection of stars within 100 parsecs was chosen, where reddening is negligible and the main features of the HR diagram are resolved (see Fig.~\ref{fig:CMD}). 

\subsection{Spectroscopy}
\label{sec:data:spec}

Spectroscopic data was collected for 13 objects using various instruments. This mainly includes the 4.1-meter Southern Astrophysical Research (SOAR) telescope, using the blue camera of the Goodman spectrograph \citep{Goodman_2004}, with the 930 mm$^{-1}$ grating and 1 arcsec slit, resulting in a resolution $\Delta \mathrm{\lambda}\approx2.7$ across the wavelength range of 3700 to 5200~$\mathrm{\AA}$. We also use Lick Observatory's 3-meter Shane telescope, in the blue arm of the Kast double spectrograph \citep{Kast_1994}, with the 600/4310 grism and 2 arcsec slit, which provides a very similar resolution and wavelength coverage as SOAR/Goodman. For two objects, ZTF-sdBV6 and ZTF-sdBV8, we acquired time-series spectroscopy using the southern Gemini Multi-Object Spectrograph (GMOS-S; \citealt{GMOS_2004}) in its long-slit mode, using the B1200 grating and a 0.75 arcsec slit, which results in a slightly higher resolution of $\Delta \mathrm{\lambda}\approx1.8$ over the same wavelength range as SOAR/Goodman and Shane/Kast. 

For each object, the instrument used to collect spectra, along with the total number of spectra used in our analysis, are listed in Table~\ref{tab:data}. The seven objects which time-series spectra were collected for are denoted by an asterisk next to the instrument. Exposure times for these observations range from 40 to 60~s for objects with pulsation periods near 6~min, and 90 to 120~s for objects with periods longer than 10~min. These short integrations ensure that amplitude reduction caused by phase smearing throughout the variability cycle is minimal. For the objects not observed in time-series, a sequence of three long exposures were taken, each covering at least one full pulsation cycle and then median combined to decrease the likelihood that errors are introduced by observing only at certain points in the pulsation phase.

Spectral images from SOAR/Goodman were CCD reduced using standard bias subtraction and flat-field division, as well cosmic ray removal, with the Goodman Spectroscopic Data Reduction Pipeline \citep{Goodman_reduce_pipeline_2020}. Due to our custom configuration, final aperture trace, background subtraction and extraction, as well as wavelength calibration, was done using the current community maintained implementation of the Image Reduction and Analysis Facility (\texttt{iraf}) \citep{iraf_1986,iraf_1993}. Data sets taken with the Shane/Kast spectrograph at Lick were CCD reduced and wavelength calibrated entirely using the \texttt{PypeIt} spectroscopic data reduction pipeline \citep{pypeit_2020}. Gemini data was reduced and wavelength calibrated with the Data Reduction for Astronomy from Gemini Observatory North and South (\texttt{DRAGONS}) reduction pipeline for the GMOS-S instrument \citep{dragons_2019}.

\section{Analysis}
\label{sec:analysis}
\begin{figure*}
\centering
\begin{minipage}[h!]{0.50\textwidth}
    \centering
    \includegraphics[width=\textwidth]{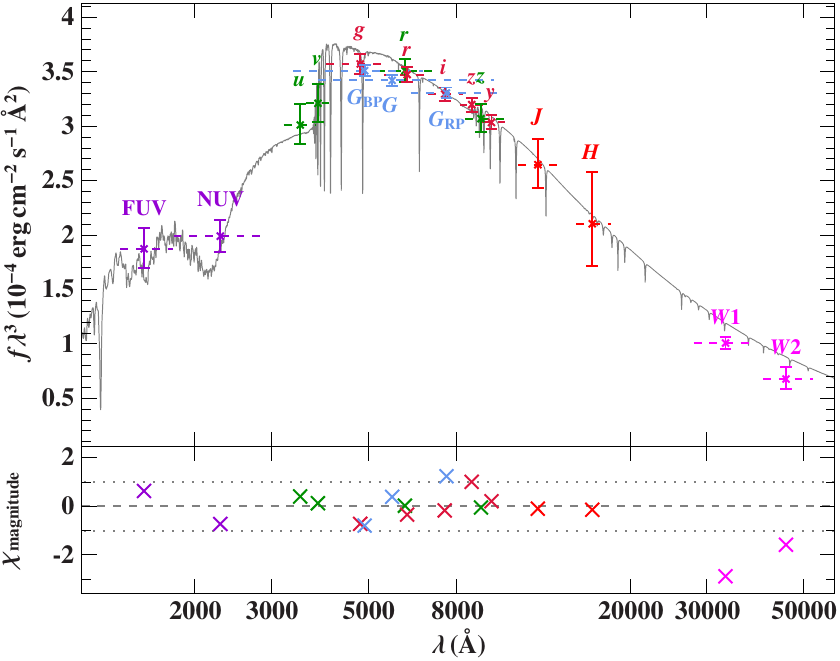}
\end{minipage}%
\begin{minipage}[h!]{0.50\textwidth}
    \centering
    \includegraphics[width=\textwidth]{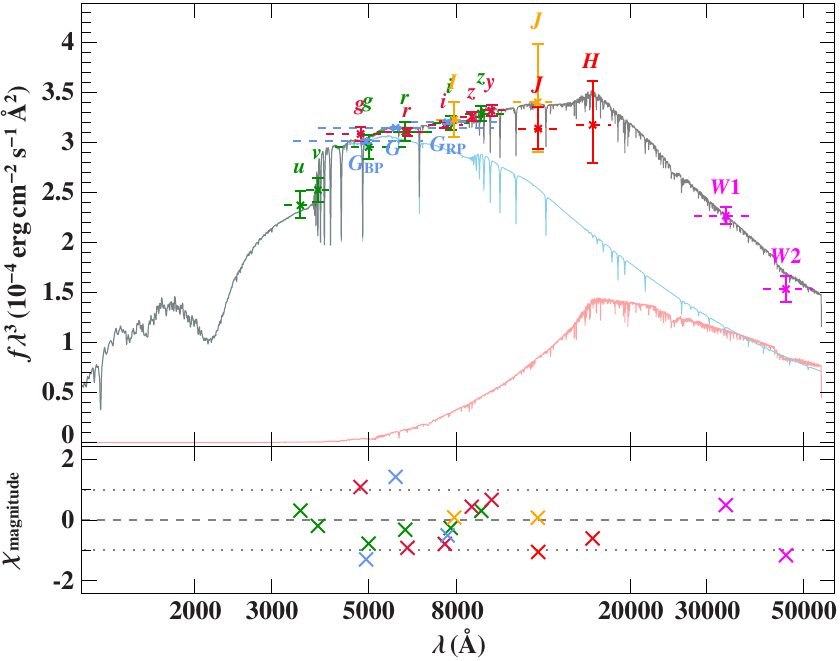}
\end{minipage}
\vspace{1mm} 
\caption{\textit{Left panel}: SED fitting for a single sdB (ZTF-sdBV1) where the {gray line} represents the best fitting sdB model spectrum. \textit{Right panel}: composite SED for a sdB + cool companion (ZTF-sdBV6) where the blue line denotes the sdB model and the red line shows the cool companion model from the PHOENIX grid, while the gray line is the composite spectrum. Colored points on both panels represent photometric flux measurements from a variety of sources, which are listed in Appendix~\ref{sec:apped:sed}.}
\label{fig:example_SEDs}
\end{figure*}
\subsection{Photometry}
\label{sec:analysis:phot}

In order to determine a precise pulsation period of our sample and search for any additional periodicities evident in the extended photometric baseline, we combined both ZTF $g$ and $r$ band data in order to extract a frequency spectrum for each star. Before creating each combined light curve, the data for both bands were individually corrected for any outliers by removing data points lying above $\pm$3$\sigma$ from the average magnitude. Once the combined $g$ and $r$ band light curves were created for all objects, we constructed each frequency spectrum by using a discrete Fourier transform (DFT) method implemented in \texttt{Period04} \citep{P04_2005}. Due to the long-term, ground-based nature of each data set, the resulting periodograms incur aliasing affects due to various observing cadences. We therefore computed an initial spectral window function to isolate and remove sidelobes corresponding to these aliases. We computed the noise spectrum of each data set and applied a 5$\sigma$ threshold to determine the significance of the resulting peaks. The individual light curves in the $g$ and $r$ bands were then fitted separately for their amplitudes occurring at the detected dominant frequency, by means of a single-harmonic, least-squares fitting. 

To place the ZTF-sdBV sample on the CMD and determine their positions relative to one another and the current population of hot subdwarfs, we first color corrected their \textit{Gaia} DR3 $G$, $G_{BP}$ and $G_{RP}$ magnitudes. We used $E(V-B)$ color excess values from \citet{Green_2015} and extinction coefficients for the corresponding filters from \citet{Casagrande_2018}. We then took the dereddened $G_{mag}$ and combined it with the distance measurement obtained from parallax via $1/~\varpi$ to compute the absolute magnitude ($M_{G_{0}}$). For ZTF-sdBVs 9, 11 and 13, the parallax precision is poor and we therefore removed these points from Fig.~\ref{fig:CMD}, but list the resulting estimates for each in Table~\ref{tab:photometry}. We calculated the color indices $(G_{BP} - G_{RP})_{0}$ and plotted the points with a color mapping that is scaled to the full range of observed photometric pulsation periods we found for our sample, which are listed in column four of Table~\ref{tab:photometry}.

\subsection{Spectroscopy}
\label{sec:analysis:spec}

To investigate RV variability, an initial fitting of the individual spectra for each set of time-series observations was performed. We fitted a pseudo-Voigt profile consisting of a Gaussian, Lorentzian and an additional polynomial term to the dominant hydrogen lines to cover continuum, line and line core of the individual absorption features using a $\chi^2$ minimization, and compared the resulting shifted wavelength to their known rest wavelengths. For SOAR/Goodman spectra, we correct for flexure drift before assessing the intrinsic variability by fitting and subtracting a 3rd order polynomial to the resulting raw RV curve. A subsequent DFT was carried out on the resulting RV measurements for all time-series spectra using \texttt{Period04}. If the resulting frequency spectrum showed a significant peak corresponding to the pulsation period found from photometry, the RV curve was phase-folded on that period and binned into either 6 or 10~bins to increase the signal-to-noise (S/N), while minimizing the effect of phase-smearing. \citet{Barlow_2010} describe a similar process in more detail for the analysis of the Goodman spectra obtained for CS\,1246. To derive the value for $\Delta$RV, we fitted a single harmonic function to the phase-binned RV measurements and took the relevant amplitude coefficient as the corresponding value. For RV variability in ZTF-sdBV7 and ZTF-sdBV12, which was tending toward a saw-tooth shape (see Fig.~\ref{fig:tsspec}), we fitted a second order Fourier term to the measurements to minimize residuals. This was not needed for any other phase-resolved results.  

To measure the atmospheric properties ({\teff}, {\logg} and {\he}) of the population, we fitted the spectroscopic data to a grid of theoretical models for sdB stars using a custom minimization method implemented with the Interactive Spectral Interpretation System (\texttt{ISIS}; \citealt{isis_2000}). This model grid is comprised based on \textsc{Atlas12} \citep{kurucz_1996} atmospheres, \textsc{Detail} \citep{Giddings_1981} departure coefficients, and \textsc{Surface} \citep{Giddings_1981} model spectra, and thus uses a hybrid local thermal equilibrium (LTE) / non-LTE approach, described in detail by \citet{przybilla_2011}. These models cover a {\teff}, {\logg} and {\he} range of 15\,000 - 55\,000 K, 4.60 - 6.60 and -5.05 to 1.005, respectively. For phase-resolved atmospheric {\teff} and {\logg} measurements, all spectra were RV corrected and co-added into 6~phase-bins based on the pulsation period, and an initial model fitting was performed on each bin individually. If no clear variability in the atmospheric parameters were found during the initial fittings, the spectra were instead median combined into a single spectrum before performing the final model fitting and measuring the atmospheric parameters. During the initial atmospheric fittings, we found that all stars showed negligible rotational broadening ($v\sin(i) \approx 0$), we therefore fixed this parameter to zero before proceeding with the final fittings. 

For each object, the final atmospheric parameters were obtained by fitting our model grid either to the phase-binned spectra (for stars with detectable phase-resolved variability) or to a median-combined spectrum (for stars without measurable variability or those with only a short sequence of long-exposure spectra). The best-fit atmospheric parameters for each spectrum or phase-bin were derived from this model-grid minimization. The final uncertainties were computed by adding in quadrature the statistical uncertainties from the fitting and an adopted systematic term of 2\% in {\teff} and 0.05 in {\logg}. For stars with viable phase-resolved measurements, mean atmospheric values and variability amplitudes were obtained by fitting a single harmonic function to the phase-binned results. The uncertainties for all harmonic-fit parameters were calculated by adding in quadrature the statistical uncertainties from the harmonic fittings and the average measurement uncertainty of the individual bins in order to account for the systematic uncertainty.

\subsection{Spectral energy distributions}
\label{sec:analysis:sed}

In order to estimate the fundamental properties of these stars, we used a spectral energy distribution (SED) fitting method to calculate their mass, radius, and luminosity. These SEDs have routinely been used to measure the properties of sdBs, and the details are further described in \citet{SED}. This method retrieves photometric flux measurements from several sources
to be used in a $\chi^{2}$ fitting to the grid of synthetic spectra mentioned in Section~\ref{sec:analysis:spec}. Because {\teff}, {\logg}, and {\he} are better constrained from
spectroscopy, these parameters were fixed to the mean spectroscopic values listed in Table~\ref{tab:spec}. This leaves only the angular diameter ($\Theta$) and color excess ($E(44-55)$) caused by interstellar extinction as free parameters. For the latter we use the extinction functions of \citet{Fitz_2019}, adopting a standard Galactic reddening parameter $R(55)=3.02$. 

The best-fit angular diameter was then combined with the \textit{Gaia} DR3 parallax to calculate the stellar radius via $R = \Theta / (2\varpi)$. The mass is then calculated using the mean spectroscopic {\logg} through $M = gR^{2}/G$, where $g$ is the surface gravity and $G$ is the gravitational constant. The luminosity is subsequently obtained from $L/L_\odot = (R/R_\odot)^2(T_\mathrm{eff}/T_{\mathrm{eff},\odot})^4$. An example of such a fit is shown in the left panel of Fig.~\ref{fig:example_SEDs}.

For ZTF-sdBV3 and ZTF-sdBV6, we find a near-infrared (NIR) and infrared (IR) flux excess that is not consistent with a single sdB fit. To rule out pollution from a nearby source, we visually inspect the corresponding 2MASS \citep{SED_2MASS}, UKIDSS \citep{SED_ukidss} and WISE \cite{SED_unwise} images for field crowding. We find no obvious nearby source of contamination and therefore proceed under the assumption that the NIR/IR excess originates from a cool companion. For these objects we perform a composite SED fitting with a cool stars grid consisting of PHOENIX models developed by \citet{phoenix_grid_2013} which have {\teff} values ranging from 2300~K to 12\,000~K, {\logg} values between 2.0 and 5.0 and a constant metallically value set to solar. A composite fitting can be seen in the right panel of Fig.~\ref{fig:example_SEDs}.

\begin{figure}
    \includegraphics[width=\linewidth]{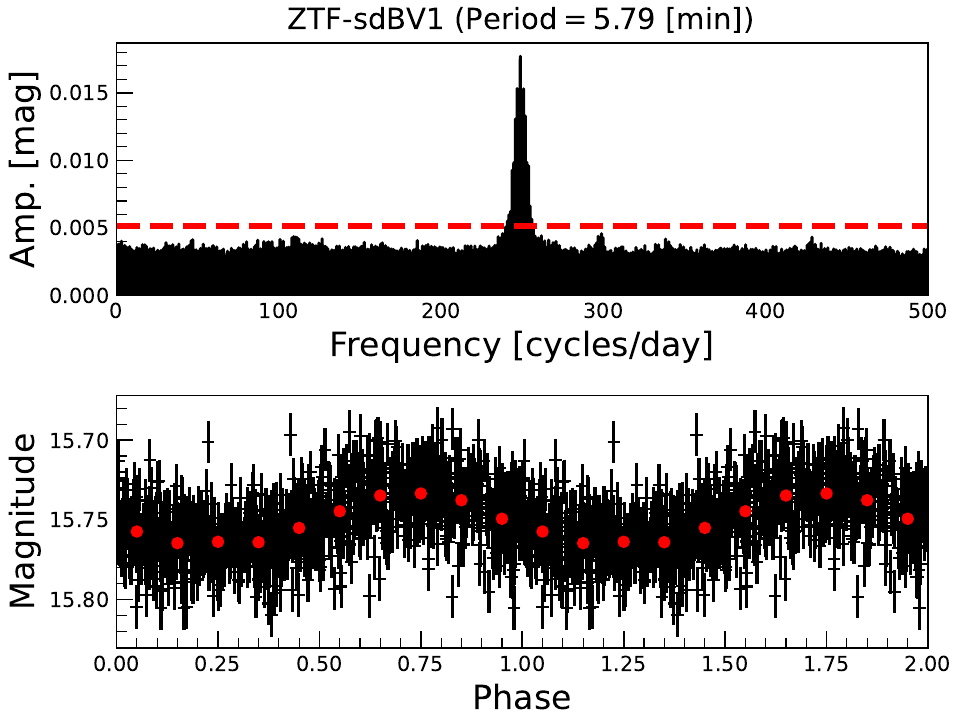}
    \caption{Photometric results for a typical rm-sdBV (ZTF-sdBV1). \textit{Top panel}: resulting DFT frequency spectrum, after removing aliasing effects, plotted along with the 5$\sigma$ detection level (red). \textit{Bottom panel}: ZTF light curve (black), phase-folded on the detected frequency, overlaid with binned values (red), and plotted over two pulsation cycles for visualization.} 
    \label{fig:sdbv1_phot}
\end{figure}

\begin{figure}
    \includegraphics[width=\linewidth]{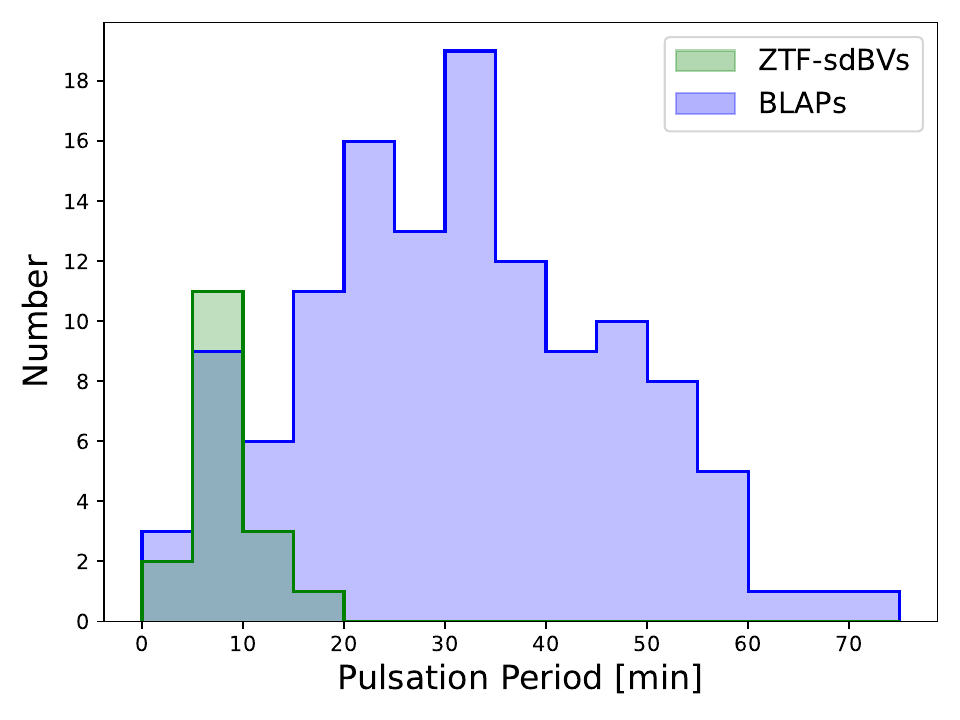}
    \includegraphics[width=\linewidth]{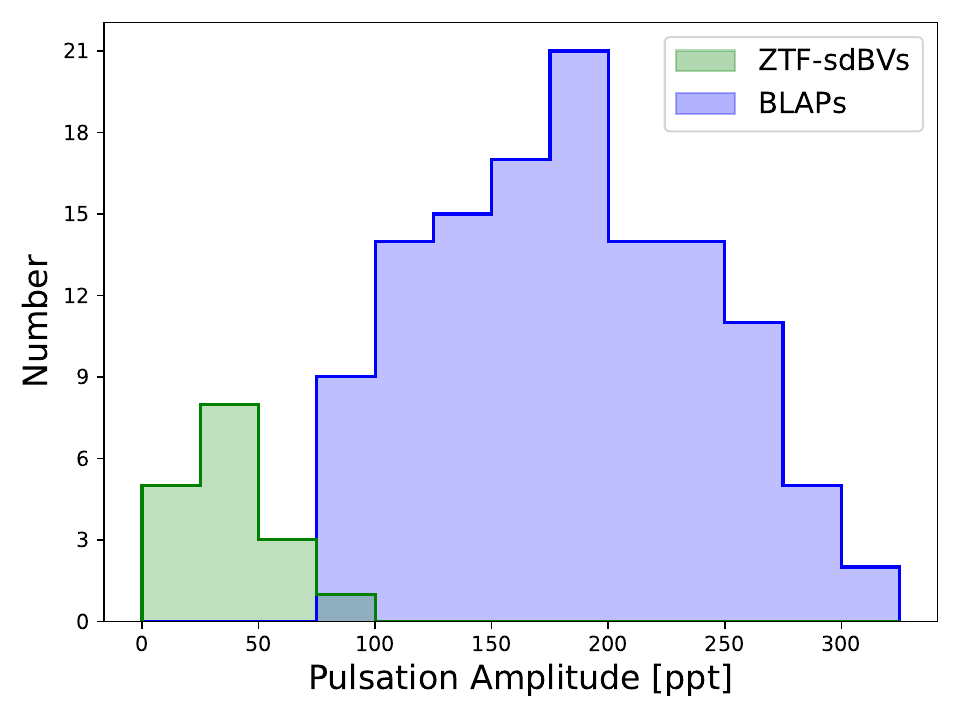}
    \caption{\textit{Top panel}: period distribution for the ZTF-sdBVs (green). For comparison, a population of BLAPs (blue) from the OGLE, ZTF and Omega White surveys are also shown. \textit{Bottom panel}: amplitude distribution for the same objects, converted from $r$ and $I$ band magnitudes to parts-per-thousand.}
    \label{fig:histograms} 
\end{figure}

\section{Results}
\label{sec:results}
\label{sec:results:spec}
\begin{figure}
    \includegraphics[width=0.95\linewidth]{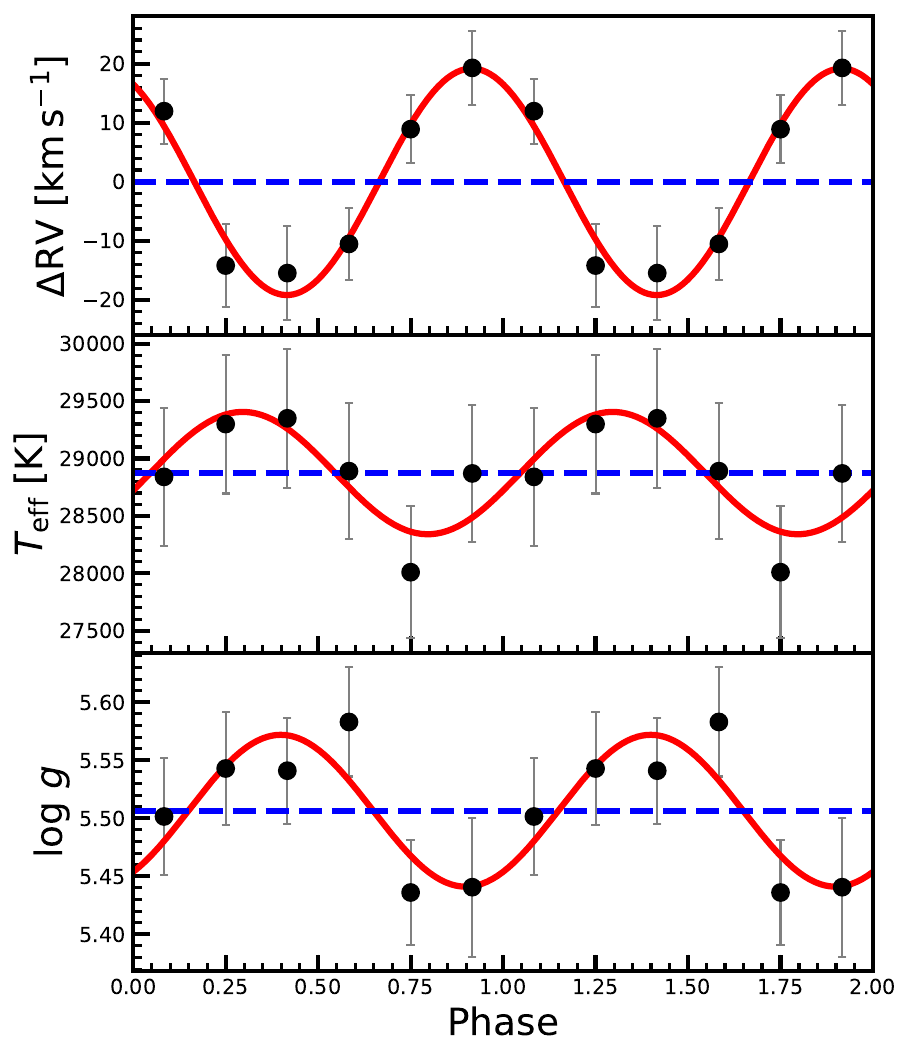}
    \caption{Phase-resolved, spectroscopic results for a typical rm-sdBV (ZTF-sdBV4). \textit{Top panel}: relative RV variations. \textit{Middle panel}: {\teff} measurements. \textit{Bottom panel}: {\logg} measurements. The measured values (black) are shown along with their single harmonic fitting (red), around the mean value (blue). All results are plotted over two pulsation cycles for visualization.} 
    \label{fig:ex_ts_spec} 
\end{figure}
\subsection{Photometry}
\label{sec:results:phot}

The ZTF-sdBVs have absolute magnitudes and color indices that place them along the EHB, where the vast majority of sdBs are found. The new objects, which were found in the \citet{gentilefusillo2021} white dwarf catalog, are also located along the EHB, providing evidence for their sdB classification. Fig.~\ref{fig:CMD} shows the CMD location for stars in our sample which have parallax uncertainties less than 25\%, color-mapped to the scale of observed pulsation periods. We see a tight clustering around a $-0.5 < (G_{BP}-G_{RP})_{0}< -0.2$ and $6<M_{G_{0}}<2$ with some objects exhibiting longer periods, deviations in absolute magnitude, or both, relative to the majority of the sample. This is a preliminary indication that these objects may differ fundamentally, which will become evident after the complete analysis.

In Table~\ref{tab:photometry}, we report all dominant frequencies found in the combined $g$ and $r$ band light curves, along with their peak-to-peak amplitudes from current ZTF photometry, which extends the baseline of that covered in \citet{ZTF_sdbv_2021}. The resulting Fourier spectrum for each can be found in the left panel of Figs.~\ref{fig:phot_grid1}~and~\ref{fig:phot_grid2} with the corresponding phase-folded $r$ band light curves plotted in the right panel of these figures. Fig.~\ref{fig:sdbv1_phot} shows the result of a typical rm-sdBV candidate (ZTF-sdBV1) in our sample with a single dominant mode from the Fourier analysis that results in sinusoidal light curve variability. The nature of the photometric variability in these stars is like that of the prototype object CS\,1246 and other radial-mode dominant sdBVs such as BA09 and BPM 36430 \citep{Barlow_2010, Oreiro_2004, Smith_2022}. Like CS\,1246, the frequency spectrum for each of these objects only reveals a single, dominant signal from the ground-based ZTF photometry. Additionally, the pulsation amplitudes for the vast majority of these stars are higher in the $g$ band, which is typical for radial-mode pulsators. We note that the $r$ band amplitudes found here (see Table~\ref{tab:photometry}) are lower than those reported in \citet{ZTF_sdbv_2021} by up to 50 mmag, depending on the object, with an average difference of 19 mmag. Potential causes for this are discussed further in Section~\ref{sec:disc}. We do not currently see evidence of any additional pulsation modes in these objects at the 5$\sigma$ sensitivity level of ZTF, which is listed for each object in Table~\ref{tab:photometry}. However, this equates to nearly a 10 mmag semi-amplitude on average. Therefore, we cannot rule out any lower-amplitude $p$ or $g$ pulsation modes given the sensitivity level of the data.

We note that several of these objects have recently been identified in the literature using the Transiting Exoplanet Survey Satellite (\textit{TESS}, \citealt{tess_2014}). This includes ZTF-sdBV7 (TIC~114196505; \citealt{Baran_2024,Uzundag_2024}) and ZTF-sdBV8 (TIC~53826859; \citealt{Krzen_2022,Baran_2024}) from the original sample presented in \citet{ZTF_sdbv_2021}. Additionally, ZTF-sdBV15 (TIC~202354658; \cite{Krzen_2022}) and ZTF-sdBV16 (TIC~968226; \citealt{Baran_2024}), from the five new ZTF-sdBVs, have been identified as sdB pulsators. Their photometric results from \textit{TESS} are discussed in Section~\ref{sec:disc}. 

 \begin{table*}[h!]
	\centering
	\caption{Photometric properties of the ZTF-sdBV sample.}
    \resizebox{\textwidth}{!}{
	\label{tab:photometry}
	   \begin{tabular}{lccccccccc}
		\hline
		Object & RA (J2000) & DEC (J2000) & P & $5\sigma$ & $A_{\textrm{ZTF}-r}$ & $A_{\textrm{ZTF}-g}$ & $G_{0}$ & $(G_{BP}-G_{RP})_{0}$ & $\Bar{\omega}$\\
         & (h: min: sec) & ($^{\circ}$:':") & (min) & (mmag) & (mmag) & (mmag) & (mag) & (mag) & (mas) \\
		\hline
		  ZTF-sdBV1 & 18:26:36.09 & 10:00:22.36 & 5.7881175(2) & 5 & $34\pm2$ & $40\pm2$ & 15.071 & -0.393 & $0.56\pm0.04$\\
		  ZTF-sdBV2 & 19:13:54.78 & -11:05:19.43 & 5.9845654(3) & 8 & $34\pm3$ & $38\pm4$ & 16.302 & -0.301 & $0.46\pm0.09$\\
        ZTF-sdBV3 & 21:03:08.51 & 44:43:14.95 & 6.097042(3) & 10 & $81\pm3$ & $95\pm4$ & 15.988 & -0.332 & $0.36\pm0.08$\\
		ZTF-sdBV4 & 08:06:07.05 & -20:08:19.07 & 6.1722840(7) & 8  & $26\pm3$ & $31\pm4$ & 16.430 & -0.338 & $0.40\pm0.05$\\
		ZTF-sdBV5 & 21:49:45.56 & 45:58:36.52 & 6.261502(4) & 7 & $32\pm3$ & $43\pm2$ & 16.645 & -0.349 & $0.31\pm0.07$\\
        ZTF-sdBV6 & 07:40:00.71 & -14:42:02.52 & 6.2630418(5) & 10 & $41\pm3$ & $57\pm6$ & 15.307 & -0.216 & $0.60\pm0.04$\\
        ZTF-sdBV7 & 19:24:08.63 & -12:58:30.83 & 6.2911382(5) & 4 & $33\pm2$ & $35\pm2$ & 14.774 & -0.402 & $0.74\pm0.04$\\
        ZTF-sdBV8 & 06:54:48.55 & -25:22:08.40 & 7.2576650(6) & 16 & $62\pm5$ & $57\pm5$ & 14.131 & -0.433 & $0.55\pm0.03$\\
        ZTF-sdBV9 & 18:52:49.08 & -18:46:08.40 & 8.801(2) & 22  & $97\pm5$ & $131\pm10$ & 17.260 & -0.485 & $0.14\pm0.15$\\
        ZTF-sdBV10 & 19:49:59.35 & 08:31:06.24 & 10.7500(3) & 11 & $33\pm4$ & $36\pm5$ & 16.249 & -0.405 & $0.39\pm0.06$\\
        ZTF-sdBV11 & 07:18:43.70 & -02:29:31.20 & 11.567145(3) & 11  & $40\pm4$ & $36\pm6$ & 17.598 & -0.356 & $0.16\pm0.13$\\
        ZTF-sdBV12 & 17:07:41.35 & -15:22:42.96 & 16.57380(2) & 10 & $57\pm3$ & $47\pm6$ & 14.160 & -0.366 & $0.44\pm0.03$\\
        ZTF-sdBV13 & 22:49:04.42 & 49:07:19.92 & 3.5125(1) & 13  & $33\pm5$ & $32\pm6$ & 17.904 & -0.419 & $0.30\pm0.11$\\
        ZTF-sdBV14 & 19:43:51.91 & 39:02:15.91 & 4.69361(2) & 7 & $27\pm2$ & $22\pm3$ & 16.726 & -0.236 & $0.56\pm0.07$\\
        ZTF-sdBV15 & 15:45:09.59 & 59:55:05.23 & 5.3928954(4) & 4 & $19\pm2$ & $16\pm2$ & 14.361 & -0.463 & $0.81\pm0.03$\\
        ZTF-sdBV16 & 21:55:17.10 & -10:25:02.59 & 5.6085396(6) & 6 & $17\pm2$ & $17\pm2$ & 14.074 & -0.439 & $1.00\pm0.04$\\
        ZTF-sdBV17 & 19:51:18.07 & 19:25:09.20 & 12.69900(5) & 7 & $19\pm3$ & $19\pm3$ & 16.201 & -0.461 & $0.60\pm0.07$\\
		\hline
	\end{tabular}}
    \tablefoot{Pulsation periods are derived from the combined $g$ and $r$ band ZTF photometry, listed along with the $5\sigma$ detection limit and corresponding peak-to-peak amplitudes for each band. The color-corrected \textit{Gaia} $G$ band magnitudes and $G_{BP}-G_{RP}$ color indices are provided, along with the parallax measurements.}
\end{table*}

The pulsations in these stars have periods ranging from 3.5 to 16.6 min, with a strong clustering near 6~min, which is coincident with CS\,1246 (6.20 min), BA09 (5.94 min), and BPM 36430 (5.69 min). The top panel of Fig.~\ref{fig:histograms} shows the period distribution of the ZTF-sdBVs compared to a sample of BLAPs from the OGLE \citep{BLAPs_2017,BLAPs_2023,BLAPs_2024,Obs_BLAPs_2024}, ZTF \citep{Kupfer_2019,ZGP_BLAP_2022}, and Omega White \citep{OW_BLAPs_2022} surveys. There is an ambiguity between the two populations for pulsation periods in the range of 5 to 10~min, making a preliminary distinction between these two classes difficult when looking for variable sdBs. On the other hand, the distribution of pulsation amplitudes creates a distinction. The bottom panel of Fig.\ref{fig:histograms} shows the peak-to-peak amplitude distribution of the ZTF-sdBVs, along with the BLAPs. A clear separation is seen between the peaks of the corresponding distributions of the two populations. The rm-sdBV candidates have pulsations amplitudes of $\approx35$~parts-per-thousand (ppt), on average. This is in contrast to the BLAPs, which typically have pulsation amplitudes of over 100 ppt. This provides initial evidence that these are two separate classes of radial-mode dominant pulsators and outlines a preliminary classification metric when analyzing time-series photometry of high-amplitude sdBVs; e.g., those with dominant pulsation mode amplitudes between 10 and 100 ppt and those with amplitudes greater than 100 ppt. However, there are exceptions to this trend and caution should be taken when when only assessing photometric variability (see Section~\ref{sec:disc}).

\subsection{Spectroscopy}

\begin{table*}
	\centering
	\caption{Atmospheric properties of the ZTF-sdBV sample.}
	\label{tab:spec}
	\begin{tabular}{lccccccccc} 
		\hline
		Object & $\Delta$RV & $T_{\textrm{eff}}$ & $\Delta T_{\textrm{eff}}$ & $\log\,g$ & $\Delta \log\,g$ & {\he}\\
         & [{\kms}] & [K] & [K] & [dex(cgs)] & [dex(cgs)] & [dex] \\
		\hline
		  ZTF-sdBV1  & $32\pm7$ & $27\,369\pm576$ & -- & $5.58\pm0.06$ & -- & $-2.93\pm0.09$ \\
		  ZTF-sdBV2  & -- & $28\,398\pm598$ & -- & $5.68\pm0.06$ & -- & $-2.95\pm0.07$ \\
        ZTF-sdBV3  & -- & $28\,477\pm709$ & -- & $5.73\pm0.08$ & -- & $-3.24\pm0.30$ \\ 
		ZTF-sdBV4  & $38\pm6$ & $28\,842\pm591$ & $1048\pm654$ & $5.51\pm0.05$ & $0.12\pm0.06$ & $-2.23\pm0.03$ \\
		ZTF-sdBV5  & -- & $27\,936\pm624$ & -- & $5.54\pm0.07$ & -- & $-2.93\pm0.11$ \\
        ZTF-sdBV6  & $18\pm3$ & $29\,115\pm602$ & -- & $5.35\pm0.07$ & -- & $-2.54\pm0.03$ \\
        ZTF-sdBV7  & $18\pm2$ & $27\,676\pm558$ & $956\pm574$ & $5.50\pm0.05$ & $0.05\pm0.05$ & $-3.29\pm0.06$ \\
        ZTF-sdBV8  & $24\pm4$ & $31\,389\pm631$ & $1249\pm759$ & $5.29\pm0.08$ & $0.12\pm0.09$ & $-1.34\pm0.01$ \\
        ZTF-sdBV9  & -- & $34\,560\pm738$ & -- & $5.70\pm0.07$ & -- & $-2.40\pm0.09$ \\
        ZTF-sdBV10  & -- & $32\,994\pm787$ & -- & $5.22\pm0.08$ & -- & $-2.94\pm0.17$ \\
        ZTF-sdBV11  & -- & $43\,549\pm961$ & -- & $5.24\pm0.11$ & -- & $-1.21\pm0.04$ \\
        ZTF-sdBV12  & $30\pm10$ & $27\,197\pm611$ & $966\pm690$ & $4.88\pm0.06$ & $0.12\pm0.06$ & $-0.71\pm0.02$ \\
        ZTF-sdBV16  & $6\pm2$ & $29\,510\pm591$ & -- & $5.63\pm0.05$ & -- & $-3.48\pm0.03$ \\ 

		\hline
	\end{tabular}
    \tablefoot{For objects with time-series spectroscopy, the average values derived from the harmonic fittings of the phase-resolved measurements are provided. The peak-to-peak amplitudes $\Delta$RV, $\Delta T_{\textrm{eff}}$ and $\Delta \log(g)$ are also listed.}
\end{table*}

\begin{figure*}
\centering
    \includegraphics[width=\textwidth]{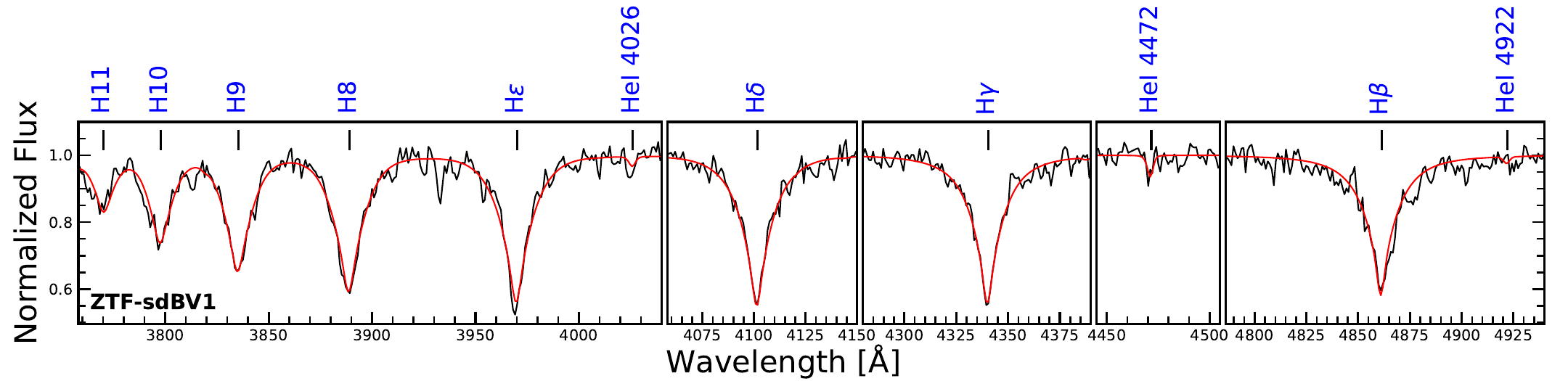}\vspace{-4.8mm}
    \includegraphics[width=\textwidth]{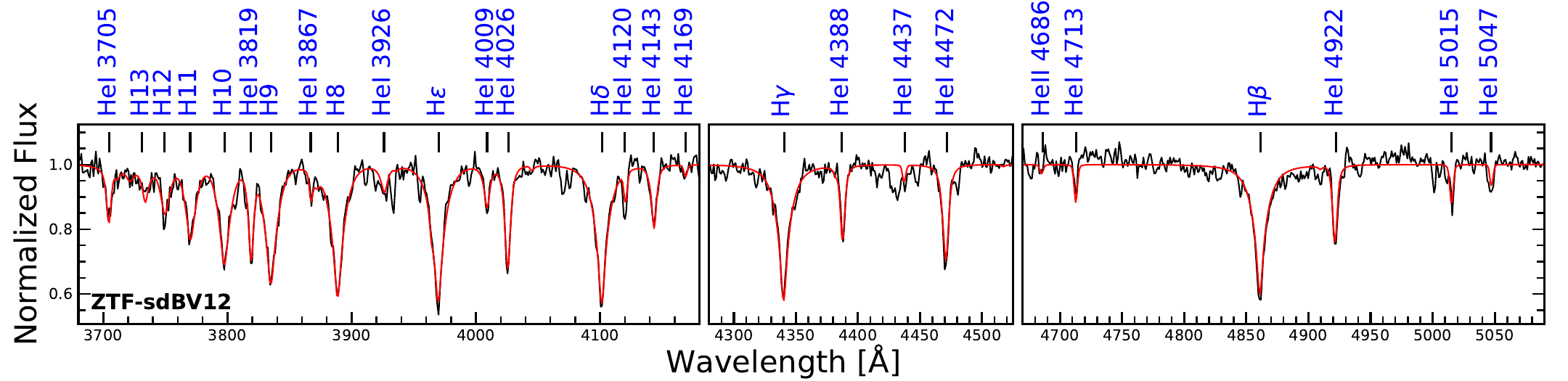}
\caption{Spectroscopic model fitting (red) to the spectra (black) collected using SOAR/Goodman for two objects in our sample. \textit{Top panel}: ZTF-sdBV1, a typical rm-sdBV spectrum. \textit{Bottom panel}: ZTF-sdBV12, a spectrum revealing a BLAP atmosphere. Notable hydrogen and helium lines are marked above each panel, in blue, at their rest wavelengths.}
\label{fig:example_spec}
\end{figure*}

In Table~\ref{tab:spec} we report the mean {\teff}, {\logg} and {\he} obtained from the spectroscopic fittings for each target, along with any corresponding variability amplitude, including RV amplitude, for the objects with time-series spectroscopy. Detectable RV variability, coincident with the photometric pulsation period, was found for all objects with time-series observations and the peak-to-peak amplitudes are reported. The average peak-to-peak $\Delta$RV that we measure is $\approx20$~{\kms}. Additionally, the typical {\teff} variability is $\approx1000$~K, on average, with corresponding {\logg} amplitudes between 0.04 and 0.12 dex. The atmospheric variability observed on the stellar surface is consistent with vertical expansion and contraction and provides evidence that the pulsations of these stars are dominated by low-order oscillations, likely being radial modes such as those found in CS\,1246, BA09 and BPM 36430. We note that the uncertainties on the fitted amplitudes are relatively large because we include an adopted systematic error term in the total uncertainty. To evaluate the significance of the variability, we computed a $\chi^{2}$ for both a harmonic model and a constant (straight-line) model through the mean. For every object, the harmonic model yields $\chi^{2}\ll1$, whereas the constant model gives a $\chi^{2}\gg1$, demonstrating that the variability is statistically significant and suggests that our uncertainties may be overestimated.

The mean atmospheric values for these stars are reported from the mean coefficient of the harmonic fittings, seen as the blue dashed line in Figs.~\ref{fig:ex_ts_spec} and \ref{fig:tsspec}. For objects in which no clear variability was determined in the phase-binned spectra, or only a sequence of long exposure spectra were available, we report the values obtained from their median combined spectra. The mean {\teff} range of our sample is $\approx27\,000 - 33\,000$~K, while {\logg} values are $\approx4.88 - 5.73$. The larger range of measurements is caused by several particular stars that are fundamentally different from the bulk of our population (see Section~\ref{sec:disc}). Fig.~\ref{fig:example_spec} shows an example fitting for two objects, ZTF-sdBV1 and ZTF-sdBV12. ZTF-sdBV1 has hydrogen dominant spectral features, typical for the rm-sdBV candidates (see Fig~\ref{fig:rmsdBV_spec_grid}). In contrast, ZTF-sdBV12 has an enhanced helium abundance and spectral features, which are more consistent with that of a BLAP \citep{Obs_BLAPs_2024}.

\begin{figure*}
\centering
\begin{minipage}[h]{0.50\textwidth}
    \centering
    \includegraphics[width=\textwidth]{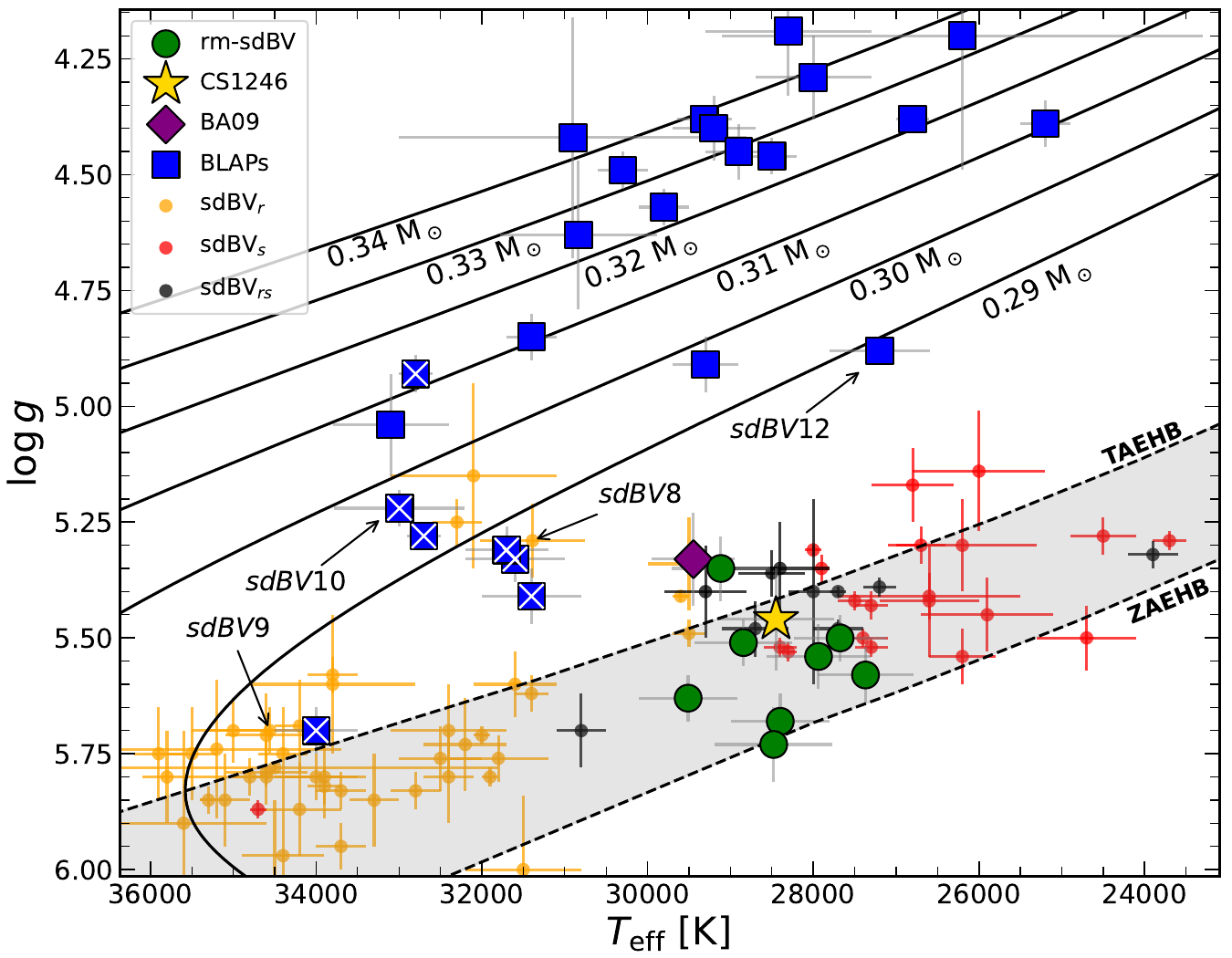}
\end{minipage}%
\begin{minipage}[h]{0.50\textwidth}
    \centering
    \includegraphics[width=\textwidth]{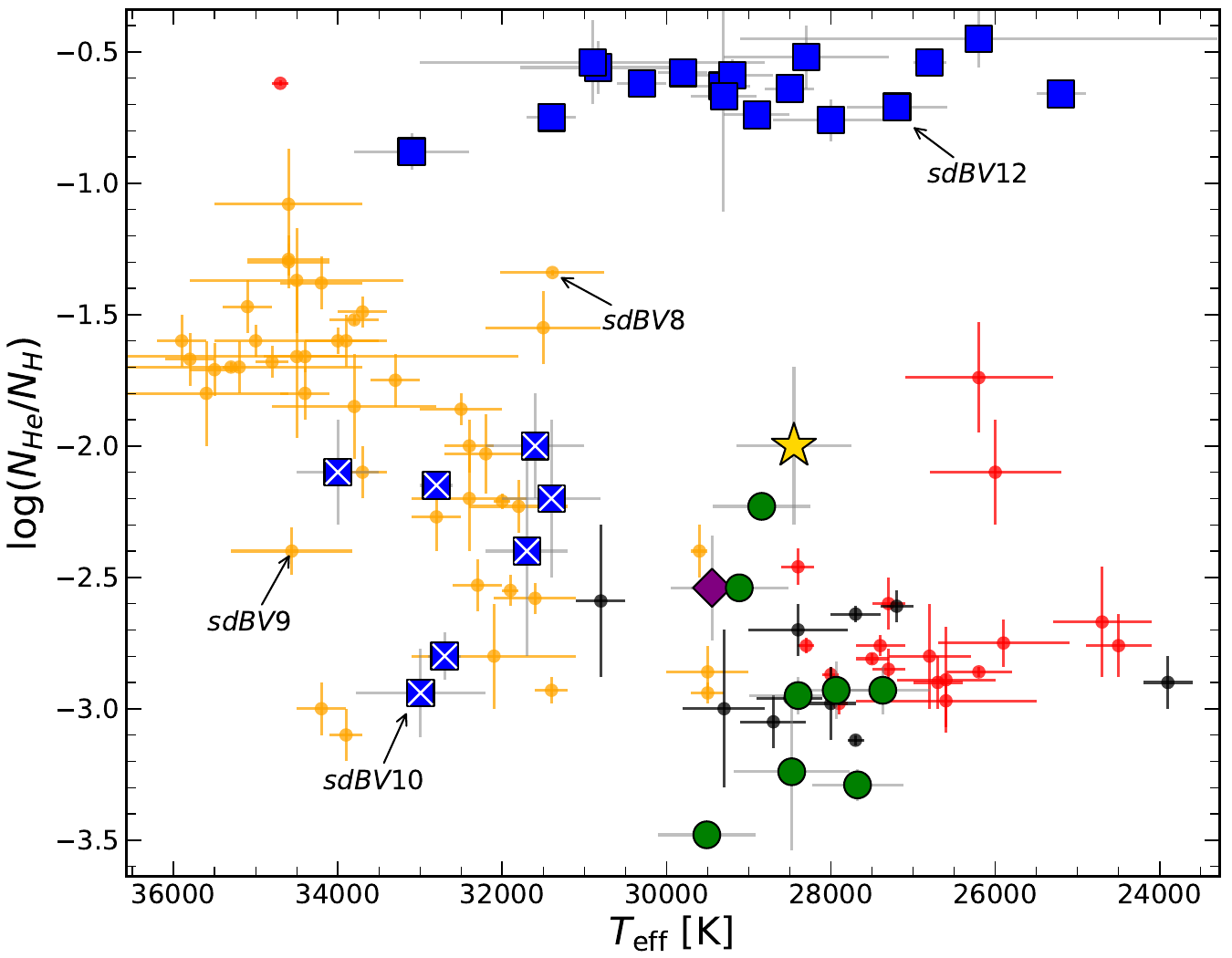}
\end{minipage}
\vspace{1mm}
\caption{\textit{Left panel:} {\teff} -- {\logg} diagram for sdB pulsators, including model tracks for the EHB (shaded gray; \citealt{EHB_1993}) and He-core pre-WDs with masses between 0.29 - 0.34~{\msol} (black lines; \citealt{Kupfer_2019}). \textit{Right panel:} {\teff} -- {\he} distribution for the same objects. The classified rm-sdBVs from ZTF are plotted with green circles along with CS\,1246 and BA09, plotted using a gold star and purple diamond, respectively. Blue squares are the known BLAPs with atmospheric properties obtained from spectra; those marked with a white X represent He-poor ({\he} < -1.5) BLAPs. Orange, red and black points represent sdBV$_{r}$, sdBV$_{s}$ and sdBV$_{rs}$ stars, respectively (sample from \citealt{sdBV_table_2017}). Arrows with labels indicating the ZTF-sdBVs classified as a BLAP or sdBV$_{r}$ are included.}
\label{fig:keil}
\end{figure*}

Fig.~\ref{fig:keil} shows that ZTF-sdBV1 - ZTF-sdBV7 and ZTF-sdBV16, cluster in the same region of the {\teff} -- {\logg} and {\teff} -- {\he} planes, where CS\,1246 and BA09 lie, making these stars consistent in both their pulsational and atmospheric properties. We therefore characterize these objects as rm-sdBVs. The remaining ZTF-sdBVs for which spectra were obtained have deviating spectroscopic and fundamental properties when compared with the majority of the sample, leading them to be classified as either a BLAP, sdBV$_{r}$ or sdOV, as noted in the last column of Table~\ref{tab:fund_props}. Their corresponding spectral fittings can be seen in Figs.~\ref{fig:sdBVr_spec},~\ref{fig:BLAP_spec} and \ref{fig:sdOV_spec}. These objects are discussed further in Section~\ref{sec:disc}. 

The atmospheric measurements of the objects which we classify as rm-sdBVs lie in the same region on the {\teff} -- {\logg} diagram as the hybrid sdBV$_{rs}$ stars, just on the high temperature boundary of the sdBV$_s$ stars as seen in Fig.~\ref{fig:keil}. This figure also shows that the rm-sdBVs lie along the EHB when compared with model tracks from \citet{EHB_1993}, and all of the BLAPs lie above the EHB and along the low-mass He-core pre-WD model tracks from \citet{Kupfer_2019}, which have been shown to accurately describe their observational and fundamental properties (see \citealt{Kupfer_2019,Bradshaw_2024,Obs_BLAPs_2024}).

\subsection{Fundamental properties}
\label{sec:results:props}

\begin{figure}
    \includegraphics[width=\linewidth]{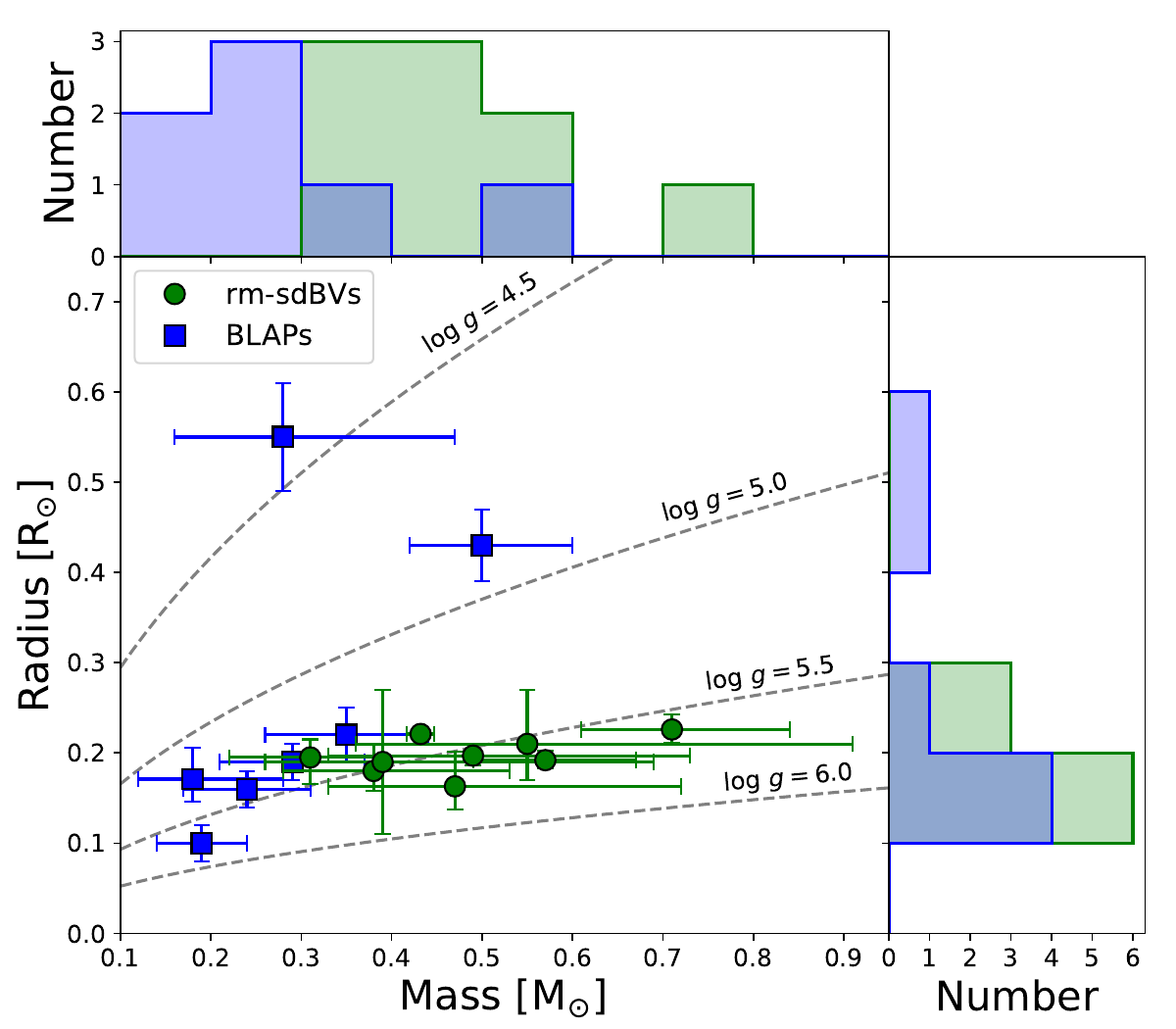} 
    \caption{Mass -- radius distribution for the rm-sdBVs (green), which includes the properties of BA09 \citep{BA09_mass_2008} and CS\,1246 \citep{Barlow_2010}. The BLAPs (blue) with available mass and radius measurements from \citet{Kupfer_2019} and \citet{Bradshaw_2024}, along with the two classified here (ZTF-sdBV10 and ZTF-sdBV12) are plotted for comparison. Lines of constant {\logg} are drawn with gray dashed lines. Histograms of the mass and radius distributions for both populations are included, above the corresponding axis, in their respective colors.} 
    \label{fig:mass_radius}
\end{figure}

We estimated the mass, radius and luminosity of this sample of objects using the SED fitting method described in Section~\ref{sec:analysis:sed}. The atmospheric values were fixed to the average measurements found from spectroscopy and the results are listed in Table~\ref{tab:fund_props}, along with a classification for each object. The reasoning for these classifications is detailed in Section~\ref{sec:disc}. For ZTF-sdBV9 and ZTF-sdBV11, we did not attempt an SED fitting since this method relies on an accurate parallax measurement, which is unreliable for these stars. The majority of masses and radii of the classified rm-sdBVs in this sample agree, within the uncertainty, with the properties expected for canonical-mass EHB stars of $\approx0.47$~{\msol} and $\approx0.20$~{\rsol}. Both CS\,1246 and BA09 have similar masses of $0.39_{-0.13}^{+0.30}$~M$_{\odot}$ and $0.432\pm0.015$~{\msol}, respectively \citep{Barlow_2010,BA09_mass_2008}. Fig.~\ref{fig:mass_radius} shows the masses and radii for the rm-sdBVs classified here, which includes the measurements of CS\,1246 and BA09. For the two stars in our sample which we classify as BLAPs, based on their derived properties, we plot them along with the mass and radius estimates which have been reported for other BLAPs \citep{Kupfer_2019,Bradshaw_2024}. We note that for two stars (ZTF-sdBV1 and ZTF-sdBV8), the masses are higher than the canonical sdB mass, although a recent study by \citet{sdB_mass_update_2024} has shown that the sdB mass range may extend higher to $\approx0.64$~{\msol}, if the sdBs have evolved from more massive progenitors. This would place both ZTF-sdBV1 and ZTF-sdBV8 inside this range, within their uncertainty.  

ZTF-sdBV10 has a lower mass than the majority of these stars. However, this mass is in agreement with the mass range of the short-period BLAPs reported by \citet{Kupfer_2019}. The mass of ZTF-sdBV12 is comparable to that of the rm-sdBV measurements but its radius is much larger, this is consistent with the lower surface gravity measurement and what has been reported for stars in the OGLE-BLAP sample \citep{Bradshaw_2024,Obs_BLAPs_2024}. We therefore classify these two stars as BLAPs, based on their spectroscopic and fundamental properties (see Section~\ref{sec:disc:BLAPs}).

In performing the SEDs for our sample we find that two objects (ZTF-sdBV3 and ZTF-sdBV6) have NIR/IR excess which does not align with a typical single sdB. For ZTF-sdBV9 and ZTF-sdBV11, we also inspected the photometry, but found no clear indication of NIR/IR excess. Fig.~\ref{fig:SED_grid2} shows that, for ZTF-sdBV3 and ZTF-sdBV6, a combined sdB + cool K/M-dwarf model spectra is an accurate fit to the photometric data. From these fittings, we calculate estimates for the companion (C$_{2}$) effective temperatures. For ZTF-sdBV3, we find a {\teff}$_{\textrm{C}_{2}}$ = $5100^{+400}_{-600}$~K, for ZTF-sdBV6, we find {\teff}$_{\textrm{C}_{2}}$ = $3900^{+1500}_{-500}$~K.

\begin{table}
	\caption{Fundamental properties of the ZTF-sdBVs.}
    \resizebox{\linewidth}{!}{
	\label{tab:fund_props}
	\begin{tabular}{lcccccr} 
		\hline
		Object & Mass & Radius & Luminosity & Class \\
         &  [{\msol}] & [{\rsol}] & [{\lsol}]\\
		\hline
        \vspace{1mm}
		  ZTF-sdBV1 & $0.71^{+0.13}_{-0.10}$ & $0.226^{+0.017}_{-0.015}$ & $26^{+05}_{-04}$ & rm-sdBV  \\
        \vspace{1mm}
		  ZTF-sdBV2 & $0.47^{+0.25}_{-0.14}$ & $0.163^{+0.038}_{-0.026}$ & $16^{+09}_{-05}$ & rm-sdBV  \\
        \vspace{1mm}
        ZTF-sdBV3 & $0.49^{+0.48}_{-0.29}$ & $0.160^{+0.070}_{-0.060}$ & $15^{+14}_{-9}$ & rm-sdBV  \\ 
        \vspace{1mm}
		ZTF-sdBV4 & $0.41^{+0.15}_{-0.10}$ & $0.180^{+0.029}_{-0.022}$ & $20^{+08}_{-05}$ & rm-sdBV \\
        \vspace{1mm}
		ZTF-sdBV5 & $0.55^{+0.36}_{-0.19}$ & $0.210^{+0.060}_{-0.040}$ & $24^{+16}_{-19}$ & rm-sdBV  \\
        \vspace{1mm}
        ZTF-sdBV6 & $0.31^{+0.08}_{-0.09}$ & $0.195^{+0.020}_{-0.029}$ & $25^{+06}_{-07}$ & rm-sdBV  \\
        \vspace{1mm}
        ZTF-sdBV7 & $0.49^{+0.24}_{-0.16}$ & $0.197^{+0.012}_{-0.011}$ & $20^{+05}_{-04}$ & rm-sdBV \\
        \vspace{1mm}
        ZTF-sdBV8 & $0.68^{+0.19}_{-0.15}$ & $0.316^{+0.021}_{-0.019}$ & $95^{+16}_{-13}$ & sdBV$_{r}$ \\
        \vspace{1mm}
        ZTF-sdBV9 & - & - & - & sdBV$_{r}$ \\
        \vspace{1mm}
        ZTF-sdBV10 & $0.18^{+0.10}_{-0.06}$ & $0.171^{+0.035}_{-0.025}$ & $31^{+15}_{-09}$ & BLAP \\
        \vspace{1mm}
        ZTF-sdBV11 & - & - & - & sdOV \\
        \vspace{1mm}
        ZTF-sdBV12 & $0.50^{+0.10}_{-0.08}$ & $0.430^{+0.04}_{-0.04}$ & $89^{+19}_{-15}$ & BLAP \\
        \vspace{1mm}
        ZTF-sdBV16 & $0.57^{+0.10}_{-0.08}$ & $0.192^{+0.010}_{-0.009}$ & $25^{+04}_{-03}$ & rm-sdBV \\

		\hline
	\end{tabular}}
    \tablefoot{Measurements are derived from the SED fitting method after fixing atmospheric values to those listed in Table~\ref{tab:spec}. The proposed classification of these objects, based on all results, is listed in the last column.}
\end{table}

\section{Discussion}
\label{sec:disc}
\subsection{Peculiar objects}

\subsubsection{BLAPs -- ZTF-sdBV10 and ZTF-sdBV12}
\label{sec:disc:BLAPs}

We classify two objects as BLAPs based on their spectroscopic and fundamental properties. ZTF-sdBV10 has a longer pulsation period, higher {\teff} and lower {\logg} than the distribution of objects denoted as rm-sdBVs. However, these values, along with its low mass measurement, are in agreement with the BLAPs, particularly the sample reported by \citet{Kupfer_2019} which have masses between 0.19 and 0.35~{\msol}. ZTF-sdBV12 has the longest pulsation period of the ZTF-sdBV sample. The atmospheric measurements for this object show a lower {\teff} and {\logg} that are in agreement with values reported for BLAPs. The helium abundance is a clear indication of a fundamental difference between this object and the sdBV classes. The spectral features of this star are very similar to those seen in BLAP spectra \citep{Bradshaw_2024,Obs_BLAPs_2024}. However, the photometric pulsation amplitudes for both of these stars are significantly lower than for the majority of currently reported BLAPs. 

Amplitude reduction in BLAPs has been previously reported in a few cases. Amplitude suppression has been attributed to light contribution from a nearby binary companion, as in the case of HD\,133729, which consists of a BLAP and a B-type main sequence star, where the BLAP was reported to have a 32~ppt peak-to-peak amplitude \citep{HD_133729_2022}. Two other cases have been observed in multi-periodic BLAPs, where amplitude modulations were found on time scales as short as a few days (ZTF J071329.02-152125.2; \citealt{koen_2024}) up to a few years (OGLE-BLAP-030; \citealt{Obs_BLAPs_2024}), resulting in peak-to-peak amplitudes as low as about 16 and 45~ppt, respectively, for the their dominant pulsation modes. We therefore apply an additional test to ZTF-sdBV10 and ZTF-sdBV12 to confirm their consistency with BLAP properties.

\citet{Obs_BLAPs_2024} found an empirical period - surface gravity relationship for BLAPs. In order to determine if the properties of these stars agree with this finding, or the relationships derived from current theoretical models, we adapt the analysis from \citet{Obs_BLAPs_2024} and plot these stars along with their derived linear {\logg} -- $\log(P)$ relationship, including theoretical models for low-mass (0.28 - 0.35~{\msol}) He-core pre-WD stars pulsating in the fundamental ($n=0$) and first-overtone ($n=1$) radial modes (from \citealt{Kupfer_2019}). We also include models for He-shell burning stars with fundamental radial-mode pulsations and masses between 0.50 and 0.80~{\msol} (from \citealt{Lin_2023}). Fig.~\ref{fig:p_vs_g} shows that both of these stars agree with these relationships, with ZTF-sdBV10 agreeing with the $n=0$ He-core-pre WD model and ZTF-sdBV12 matching with either the $n=0$, He-core-pre WD model or the $n=0$, He-shell burning model. In fact, the mass measurement for ZTF-sdBV12 is in agreement with the 0.50~{\msol} He-shell burning model. Given these results, we proceed to classify these stars as BLAPs and place them on Figs.~\ref{fig:keil}~and~\ref{fig:mass_radius} as part of the BLAP sample. 

\subsubsection{sdBV$_{r}$ -- ZTF-sdBV8 and ZTF-sdBV9}
\label{sec:disc:sdbvr}

ZTF-sdBV8 was initially found in ZTF and reported as a sdBV by \citet{ZTF_sdbv_2021} but has since been rediscovered in \textit{TESS} as a pulsating sdB by \citet{Krzen_2022} and \citet{Baran_2023}. \citet{Baran_2023} has shown, using data from \textit{TESS}, that this star has at least 11 frequencies in the sdBV $p$-mode regime. From the ZTF photometry, only one clear mode is found as a result of the DFT. This further emphasizes the need for high S/N, high-cadence observations in these objects to rule out lower-amplitude modes. We see that the atmospheric and fundamental properties of this star are different than the rm-sdBVs, with a higher {\teff} and {\he}, and a lower {\logg}. This places it near the sdBV$_{r}$ stars and short period BLAPs on the {\teff} -- {\logg} diagram. Based on these results, and those of the frequency analysis of \citet{Baran_2023}, we confirm the classification as a $p$-mode sdBV$_{r}$. We suggest that the dominant mode may be radial based on the atmospheric variability. However, more work is needed to constrain the specific degree of oscillation of this and the lower-amplitude modes. Interestingly, the helium abundance value of -1.34 lies somewhere between the sdBV$_{r}$ and BLAP clusters, with a luminosity that is significantly higher than the typical sdB, making this an unusual object for a member of this class.

ZTF-sdBV9 has atmospheric properties that coincide with the majority of the sdBV$_{r}$ class members. However, the pulsation period is longer than most sdBV$_{r}$ stars at almost 9~min and the amplitude is larger, near 100 ppt. Fig.~\ref{fig:keil} shows that it is also located next to a short period BLAP, within the sdBV$_{r}$ cluster. However, this BLAP (ZTF J071329.02-152125.2) from \citet{Kupfer_2019} has a much shorter pulsation period of 3.34~min. Fig.~\ref{fig:p_vs_g} makes it apparent that ZTF-sdBV9 does not agree with the empirical period -- surface gravity relationship derived for the BLAPs (Fig.~\ref{fig:p_vs_g}) and we therefore classify this object as a $p$-mode sdBV$_{r}$. However, we do not have time-series spectra, or an estimation of the fundamental parameters. The abnormal properties that both ZTF-sdBV8 and ZTF-sdBV9 have in comparison with the sdBV$_{r}$ class makes them interesting candidates for further investigations.

\subsubsection{sdOV -- ZTF-sdBV11}
\label{sec:disc:sdov}

ZTF-sdBV11 has very peculiar results compared to the rest of the objects in our study. Whereas the pulsation period and amplitude of this star are similar to some sdBVs, the spectroscopic properties deviate substantially from those that show consistency with sdBs. This star has a much higher {\teff} than the rest of the objects at almost 44\,000~K. The spectral fitting for this star, seen in Fig.~\ref{fig:sdOV_spec}, shows significantly stronger \ion{He}{ii} features compared to its \ion{He}{i} lines. This is a distinguishing characteristic for sdO stars \citep{Heber2016}. We therefore classify this star as a variable sdO.

\subsection{rm-sdBVs}

The objects characterized here as rm-sdBVs have similar pulsation periods as the short period BLAPs but typically smaller pulsation amplitudes. The pulsation amplitudes for the rm-sdBVs lie below 100 ppt and are lower in the $r$ band than the $g$ band. When extending the baseline of previously analyzed ZTF data, we find that the $r$ band amplitudes for the rm-sdBVs are up to 40 mmag (ZTF-sdBV5) smaller than those initially reported. There are likely two reasons for this observation. The first being that the observing cadence of the ZTF photometry for our sample has decreased over the years since their initial discovery, potentially suppressing the measured variability. The second, systematic reason, for this lower amplitude is that the objects may be experiencing amplitude modulations caused by perturbations in the driving force behind the pulsations. This behavior in sdBVs is not unprecedented, as \citet{Kilkenny_2010} has shown, sdBVs can experience amplitude variations on timescales of days to years, with some even experiencing "transient" frequencies that appear during one observing epoch and are then undetectable in others. In fact, this amplitude modulation phenomenon has been well documented in the rm-sdBV, CS\,1246 where the pulsation amplitude has experienced an exponential decay from 24 ppt to about 4 ppt over an 8~yr period, nearly reducing the radial-mode oscillation to undetectable levels \citep{Hutchens_2017}.

When evaluating where the rm-sdBVs stand spectroscopically in comparison to the other known sdB pulsators, their atmospheric properties are consistent with the $p$ and $g$-mode hybrid sdBV$_{rs}$ stars. When assessing only the classified rm-sdBVs in our sample, we find that they have a {\teff} of 28\,300~K and {\logg} of 5.56, on average. Fig.~\ref{fig:keil} exemplifies this point well in showing that they are also situated at the high temperature boundary region of sdBV$_{s}$ stars in both {\teff} -- {\logg} and {\teff} -- {\he} space when compared with a background sample of known sdBVs from \citet{sdBV_table_2017}. This separates them from the sdBV$_{r}$ stars, creating a gap around 30\,000~K between the two classes. Despite the spectroscopic similarities with the hybrid sdBV$_{rs}$ class, we see no additional pulsation modes for the rm-sdBVs in our sample above the sensitivity threshold of the ground-based ZTF photometry, as the cases of the radial-mode dominant sdBV$_{rs}$ stars BA09 and BPM 36430 have shown \citep{Oreiro_2004,Smith_2022}. Long-term, space-based observations with facilities such as \textit{TESS} are crucial to determine the full frequency spectra for these rm-sdBVs. 

Currently, two objects which we characterize as rm-sdBVs have frequency spectra reported from \textit{TESS} photometry. ZTF-sdBV7 (TIC 114196505) was reported by \citet{Baran_2024} to have a dominant mode matching the frequency reported here, along with one tentative $g$-mode frequency at a period of about 2.6 hours and amplitude of 7.3~ppt. However, this signal was stated to be just above their S/N detection threshold of 4.5, requiring further confirmation. ZTF-sdBV16 (TIC 968226) is also reported in the same study to have a higher frequency $p$-mode at 3 min with an amplitude of 2.51~ppt, in addition to the dominant mode reported here. \citet{Baran_2024} also reports 8~$p$-mode frequencies for the rm-sdBV candidate ZTF-sdBV15 (TIC 202354658), all of which closely spaced and preliminarily interpreted as pairs from rotational splitting or a binary orbit. 

\begin{figure}
    \includegraphics[width=\linewidth]{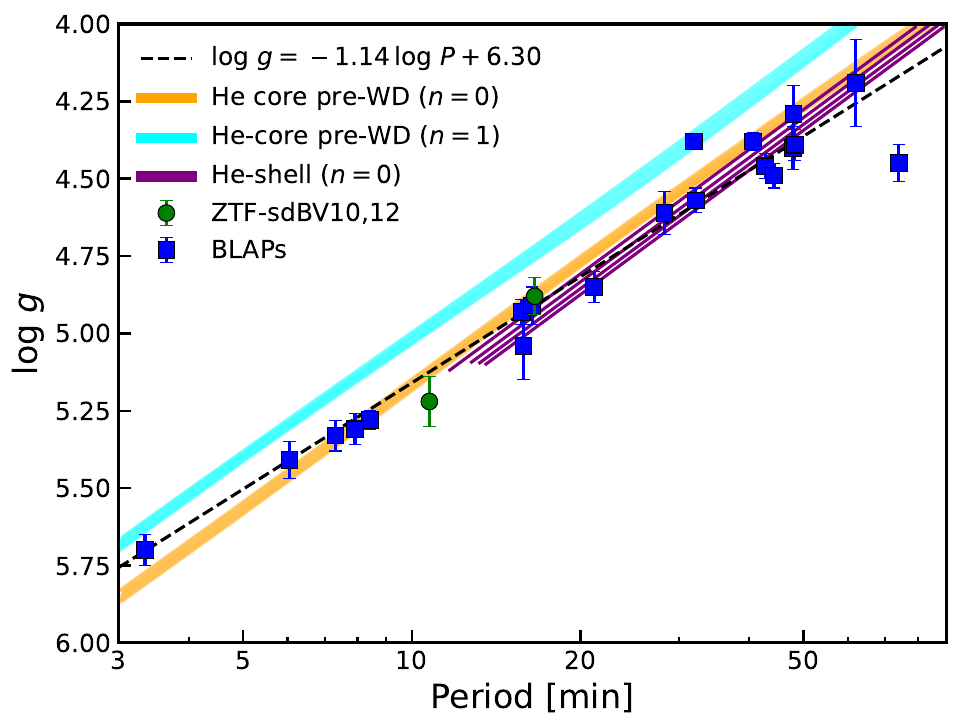}
    \caption{Pulsation period -- surface gravity relationship for BLAPs. The dashed black line represents the empirical relationship found by \citet{Obs_BLAPs_2024}. The orange and blue shaded regions are the model relationships for He-core pre-WDs in both the fundamental ($n=0$) and first-overtone ($n=1$) radial modes, respectively \citep{Kupfer_2019}. The purple lines denote the relationship for He-shell burning stars pulsating in the fundamental ($n=0$) radial mode from \citealt{Lin_2023}. The measured values for the two BLAPs, ZTF-sdBV10 and ZTF-sdBV12 (green), which we classify here are plotted along with a sample of BLAPs (blue) that have reliable spectroscopic properties reported (see \cite{Obs_BLAPs_2024}).} 
    \label{fig:p_vs_g}
\end{figure}

Fig.~\ref{fig:mass_radius} shows that the mass distribution for the rm-sdBVs is relatively well constrained, with most objects lying between 0.40 and 0.60~{\msol}. We compare these measurements to those of BLAPs with available mass and radius determinations, including the two objects classified here as BLAPs. On average, the BLAP mass distribution is shifted to lower values, with most BLAPs having masses between 0.15 and 0.40~{\msol}. As shown in Fig.~\ref{fig:mass_radius}, the BLAPs and rm-sdBVs occupy distinct regions of the mass–radius plane, implying different mean stellar densities exist between the two classes. The differences in observed pulsation properties likely arise because the two classes differ in internal structure, evolutionary state, or radial order. These effects would influence relationships derived from period mean-density arguments, apparent in the empirical {\logg} -- $\log(P)$ relationship found by \citet{Obs_BLAPs_2024} for the BLAPs. Fig.~\ref{fig:p_vs_g} shows that a BLAP with a 6 min pulsation period would have a {\logg} of $\approx5.40$, which is lower than that of the average rm-sdBV found here, which is {\logg} = 5.56. This could be explained if most BLAPs are pulsating pre-low-mass white dwarf stars (see \citealt{Romero_2018, Byrne_2018, Byrne_2020}), and the rm-sdBVs are He-core burning sdBs. 

The proposed binary formation channels for sdBs implies that the the rm-sdBVs share similar evolutionary pathways. Due to their intrinsic variability in both photometric and spectroscopic properties, traditional binary detection methods are made more difficult. However, photometric pulse timing analysis has become very useful in detecting binary companions in these stars by observing periodic light-travel-time effects (LTTEs). In fact, both CS\,1246 and BPM 36430 have had a binary companion confirmed through this approach \citep{Brad_OC_2011,Smith_2022}. The importance of this analysis in these stars is emphasized due to the fact that NIR/IR excess is seen in some of the rm-sdBVs in our sample. Confirming additional binary companions to these stars would provide key insights into their particular evolutionary pathways and provide further evidence that they also share a common evolutionary origin.

The results of this study show that these short-period, large-amplitude sdBVs comprise a group of pulsators that show distinguishable characteristics when compared with BLAPs and instead share properties with canonical sdBs. The photometric and spectroscopic variability indicates that the pulsations are caused by low-order oscillations, with characteristics similar to those of other sdBVs whose dominant mode is radial. We therefore classify these stars as rm-sdBVs. However, more detailed analysis is required to fully constrain the specific order of their dominant pulsation modes 

\section{Summary and conclusion}
\label{sec:conc}

Radial-mode pulsations in sdBs have recently become a topic of interest among variable hot subdwarfs, as more objects are being discovered which show these short-period, large-amplitude oscillations. However, broad spectroscopic measurements for these stars, which would allow for the determination of their characteristics as a whole, has been lacking. In this study, we measure and define the spectroscopic and fundamental properties of a sample of previously reported, radial-mode dominant sdBVs found in ZTF, in order to classify their standing among other known classes of sdB pulsators and BLAPs.

We show that the rm-sdBVs constitute a group of pulsating hot subdwarfs that all show very similar photometric and spectroscopic properties. These stars are consistent in terms of their photometric pulsation periods and amplitudes, as well as average atmospheric properties ({\teff} and {\logg}) and corresponding variability caused by radial-mode pulsations. The rm-sdBVs all have pulsation periods near 6~min with photometric amplitudes below 100~ppt. All objects with phase-resolved spectroscopic results show radial velocity amplitudes up to $\Delta\textrm{RV}\approx40~\textrm{{\kms}}$, effective temperature variations $\Delta T_{\textrm{eff}}\approx1000$\,K and surface gravity variations $\Delta\log\,g\approx0.1$, over their pulsation cycles. 

The mean {\teff} of the rm-sdBVs ranges between 27\,000 and 29\,000~K, with mean {\logg} values between 5.35 and 5.73~dex. This places them in the same region of the {\teff} -- {\logg} diagram as the hybrid $p$ and $g$-mode sdBV$_{rs}$ stars, leading to important questions on the commonality between these two classes. Additional investigations are ongoing to determine the extent of this similarity as well as the long-term pulsation behavior and binary nature of the rm-sdBVs.

Using their measured spectroscopic parameters combined with their distances, we fitted their SEDs to derive fundamental properties for our sample. The masses of the majority of these objects lie between 0.40 and 0.60~{\msol} with radii spanning between 0.10 and 0.30~{\rsol}, near the canonical sdB properties. These results confirm that the rm-sdBVs are distinct from BLAPs, despite both exhibiting large-amplitude radial-mode pulsations and sharing similar photometric properties.

\begin{acknowledgements}

This research was supported by Deutsche Forschungsgemeinschaft  (DFG, German Research Foundation) under Germany’s Excellence Strategy - EXC 2121 "Quantum Universe" – 390833306. Co-funded by the European Union (ERC, CompactBINARIES, 101078773). Views and opinions expressed are however those of the author(s) only and do not necessarily reflect those of the European Union or the European Research Council. Neither the European Union nor the granting authority can be held responsible for them.

MD was supported by the German Aerospace Center (DLR)
through grant 50-OR-2304. 

Based on observations obtained at the international Gemini Observatory, program ID GS-2020B-FT-202, a program of NSF NOIRLab, which is managed by the Association of Universities for Research in Astronomy (AURA) under a cooperative agreement with the U.S. National Science Foundation on behalf of the Gemini Observatory partnership: the U.S. National Science Foundation (United States), National Research Council (Canada), Agencia Nacional de Investigaci\'{o}n y Desarrollo (Chile), Ministerio de Ciencia, Tecnolog\'{i}a e Innovaci\'{o}n (Argentina), Minist\'{e}rio da Ci\^{e}ncia, Tecnologia, Inova\c{c}\~{o}es e Comunica\c{c}\~{o}es (Brazil), and Korea Astronomy and Space Science Institute (Republic of Korea).

Based on observations obtained at the Southern Astrophysical Research (SOAR) telescope, which is a joint project of the Minist\'{e}rio da Ci\^{e}ncia, Tecnologia e Inova\c{c}\~{o}es (MCTI/LNA) do Brasil, the US National Science Foundation’s NOIRLab, the University of North Carolina at Chapel Hill (UNC), and Michigan State University (MSU).

We acknowledge the use of the C. Donald Shane 3-meter telescope at the Lick Observatory, University of California Observatories for this research.

Based on observations obtained with the Samuel Oschin Telescope 48-inch and the 60-inch Telescope at the Palomar Observatory as part of the Zwicky Transient Facility project. ZTF is supported by the National Science Foundation under Grants No. AST-1440341 and AST-2034437 and a collaboration including current partners Caltech, IPAC, the Oskar Klein Center at Stockholm University, the University of Maryland, University of California, Berkeley , the University of Wisconsin at Milwaukee, University of Warwick, Ruhr University, Cornell University, Northwestern University and Drexel University. Operations are conducted by COO, IPAC, and UW.

The authors acknowledge the High Performance Computing Center (HPCC) at Texas Tech University for providing computational resources that have contributed to the research results reported within this paper. URL: http://www.hpcc.ttu.edu

\end{acknowledgements}

\bibliographystyle{aa}
\bibliography{refs}

\begin{appendix}
\section{Photometry}
\label{sec:apped:photometry}

Results from the analysis of ZTF photometry for each object in the current sample, including the frequency spectrum and the phase-folded light curve for each, where phase zero represents the start of the observation baseline for each data set.

\begin{figure*}[h!]
\centering
    \includegraphics[height=\textheight - 1.5cm]{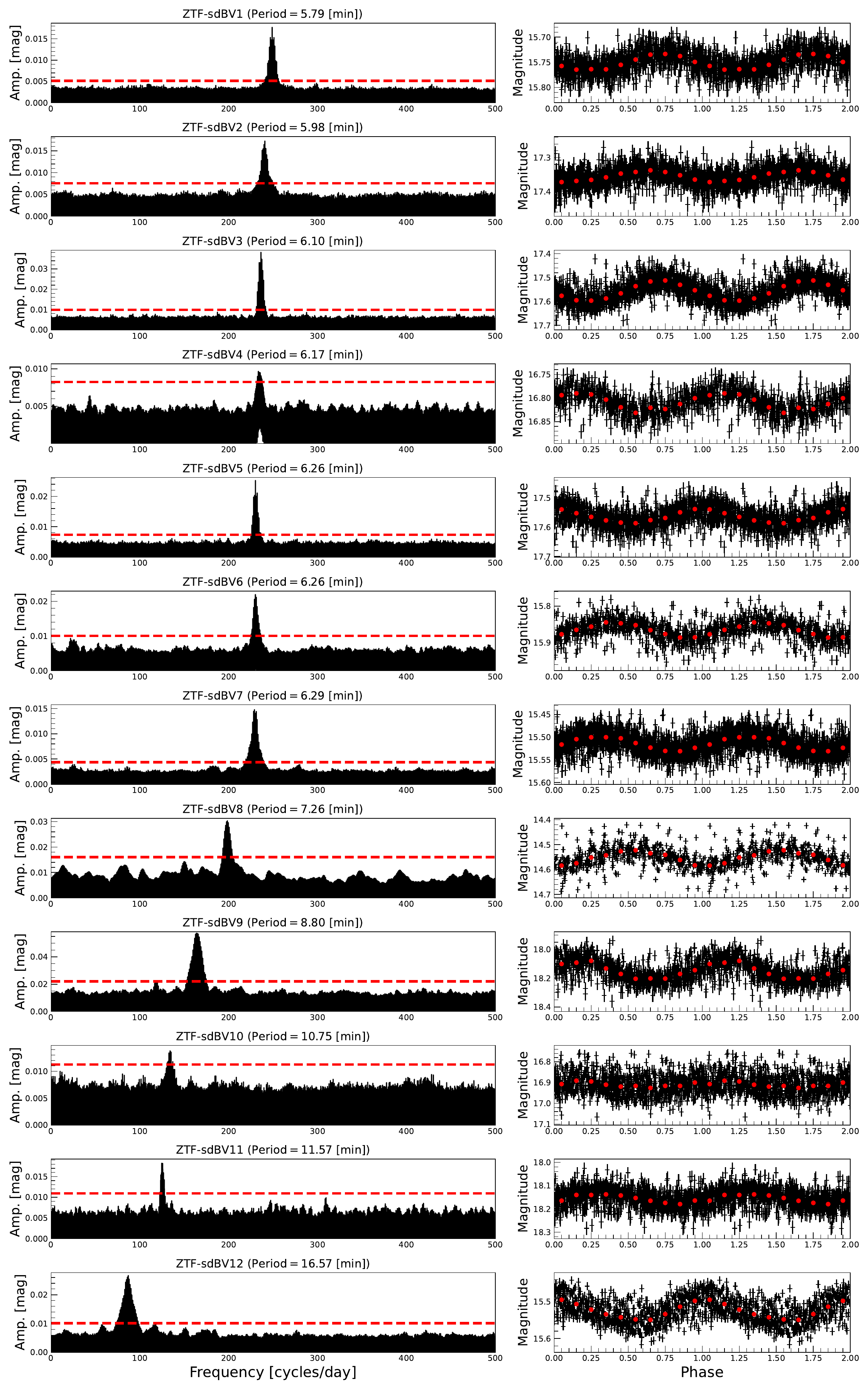}
    \caption{ZTF photometry for the original 12 ZTF-sdBVs. \textit{Left panel:} frequency spectra of the combined $g$ and $r$ band data after removing any aliases found in the spectral window function, where the dashed red line represents the 5$\sigma$ detection level. \textit{Right panel:} $r$ band light curves (black) phase-folded on the respective frequency, overlaid with binned values (red), and plotted over two pulsation
    cycles for visualization.}
\label{fig:phot_grid1}
\end{figure*}

\begin{figure*}
\centering
    \includegraphics[width=0.785\textwidth]{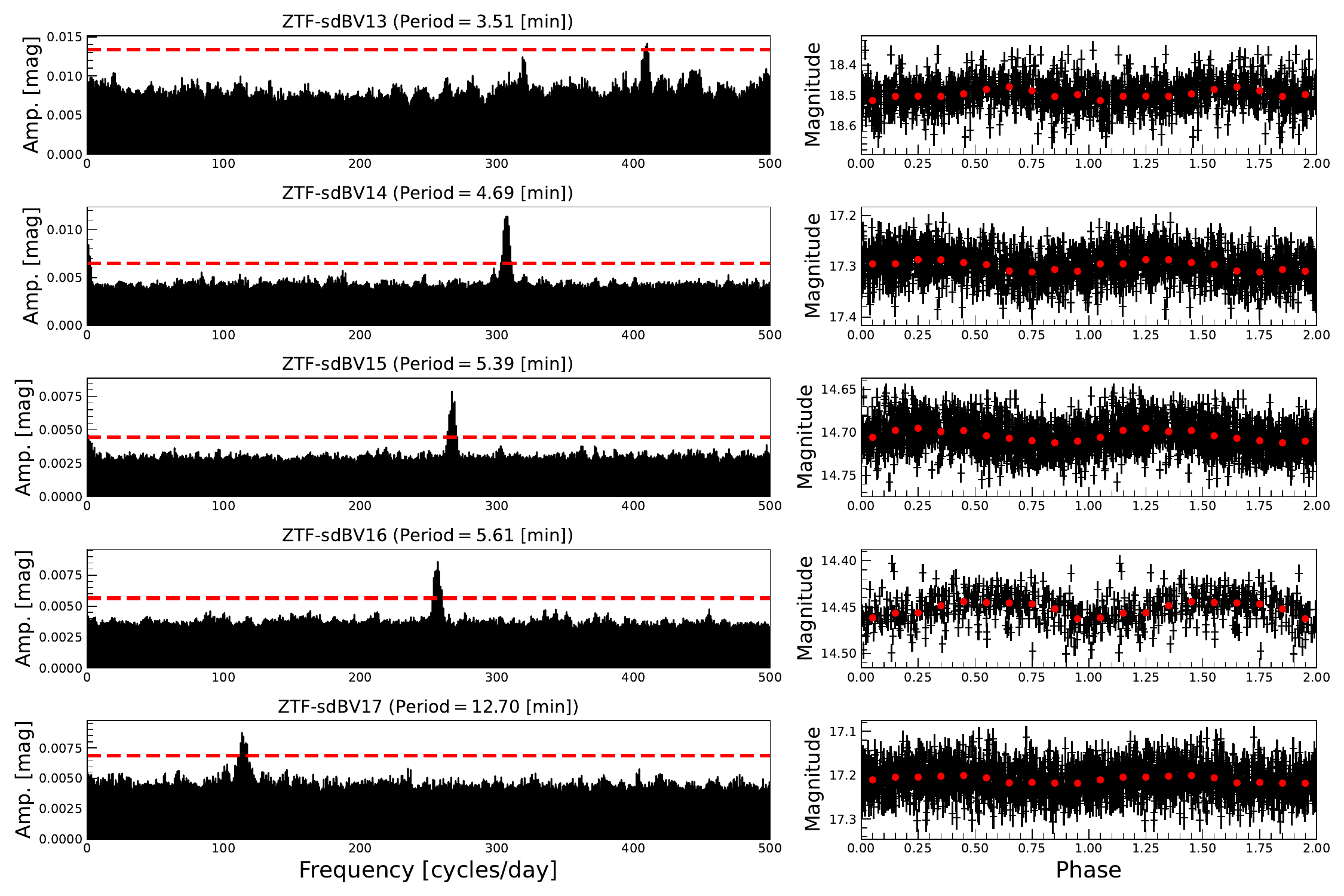}
    \caption{ZTF photometry for the five new ZTF-sdBVs. \textit{Left panel:} frequency spectra of the combined $g$ and $r$ band data after removing any aliases found in the spectral window function, where the dashed red line represents the 5$\sigma$ detection level. \textit{Right panel:} $r$ band light curves (black) phase-folded on the respective frequency, overlaid with binned values (red), and plotted over two pulsation
    cycles for visualization.}
\label{fig:phot_grid2}
\end{figure*}

\section{Spectroscopic fittings}
\label{sec:apped:spectroscopy}
\begin{figure*}
\centering
    \includegraphics[width=\textwidth]{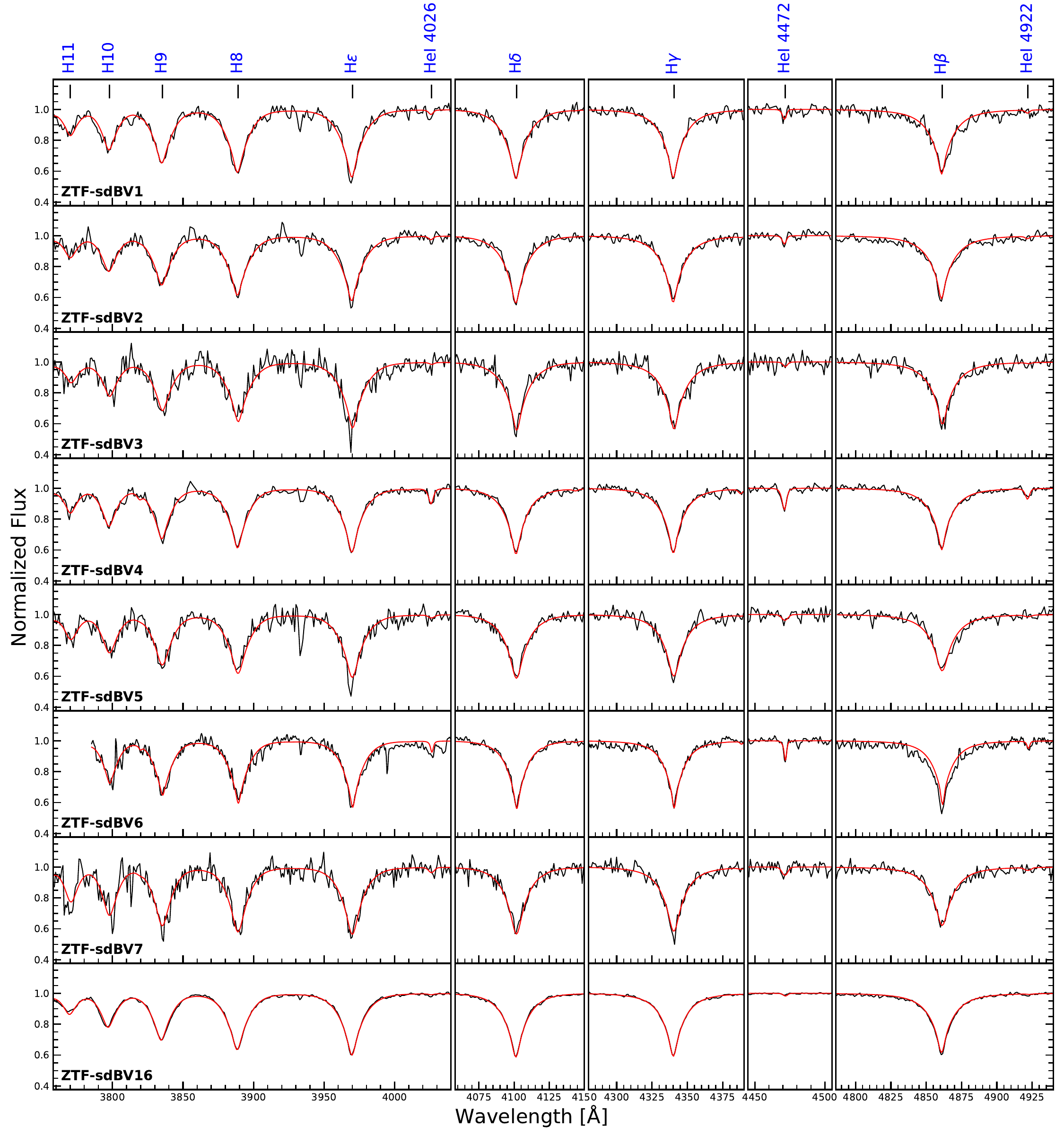}
    \caption{Spectroscopic model fitting (red) to the data (black) for rm-sdBVs classified in this sample. Objects which have phase-resolved results show the fitting for an individual bin.}
   \label{fig:rmsdBV_spec_grid}
\end{figure*}

\begin{figure*}
\centering
    \includegraphics[width=\textwidth]{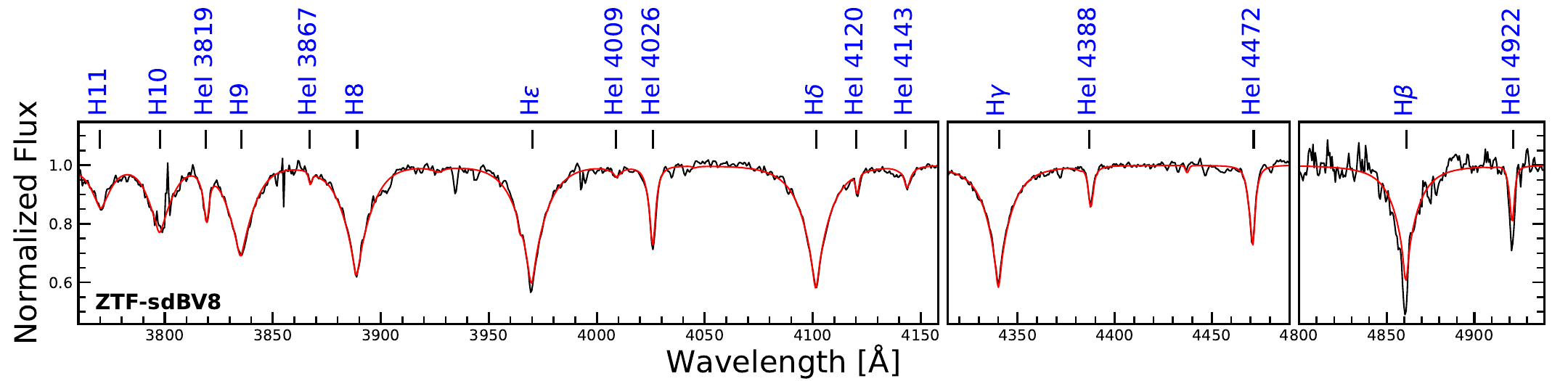}\vspace{-5mm}
    \includegraphics[width=\textwidth]{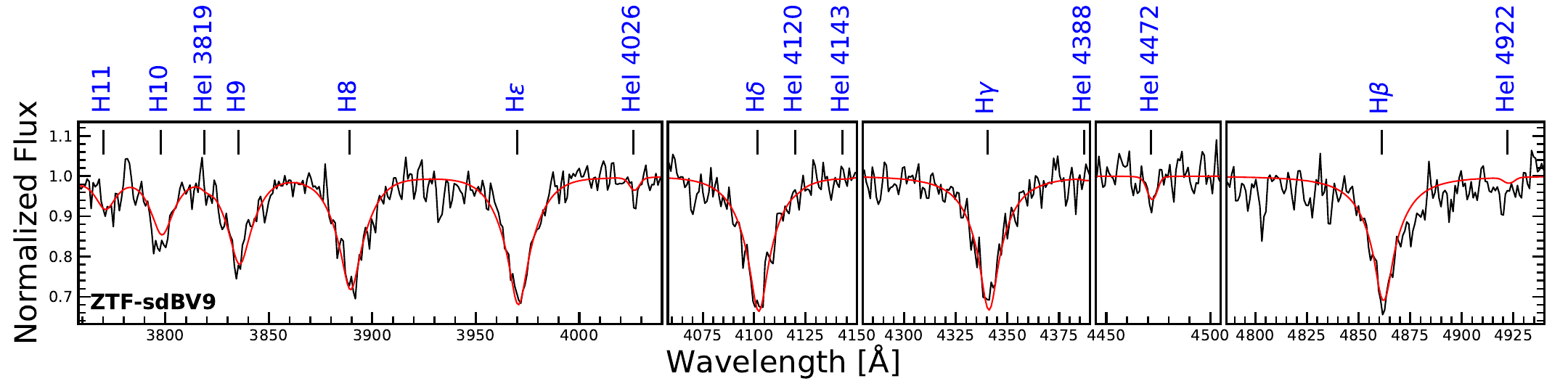}
\caption{Spectroscopic model fitting (red) to the data (black) for the sdBV$_{r}$, $p$-mode pulsators classified in this sample. \textit{Top panel:} ZTF-sdBV8. \textit{Bottom panel:} ZTF-sdBV9.}
\label{fig:sdBVr_spec}%
\centering
    \includegraphics[width=\textwidth]{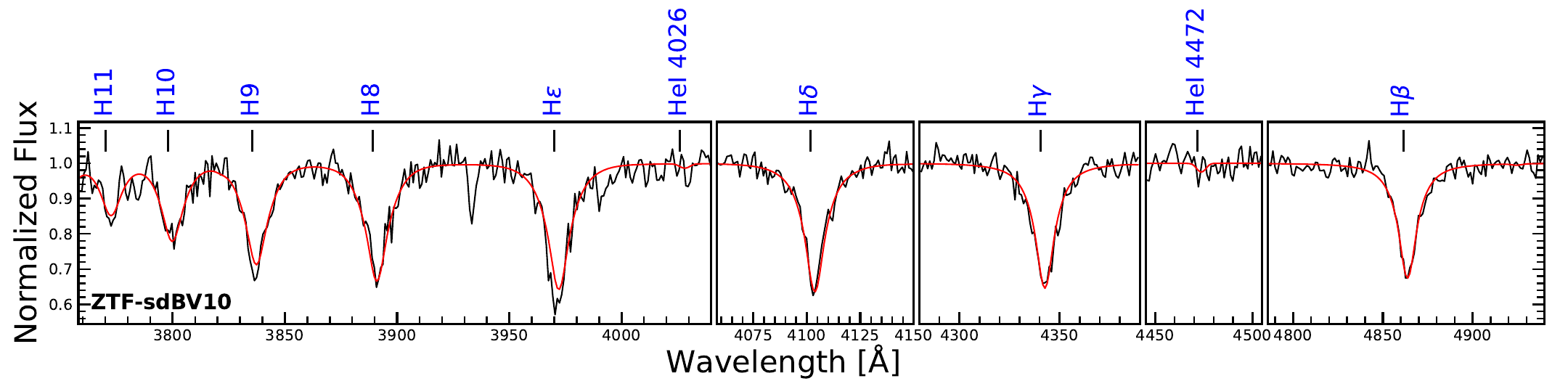}\vspace{-5mm}
    \includegraphics[width=\textwidth]{spec_figs/sdbv12_spec_fit_v2.pdf}
\caption{Spectroscopic model fitting (red) to the data (black) for the two BLAPs classified in this sample. \textit{Top panel:} ZTF-sdBV10. \textit{Bottom panel:} ZTF-sdBV12.}
\label{fig:BLAP_spec}%
\centering
    \includegraphics[width=\textwidth]{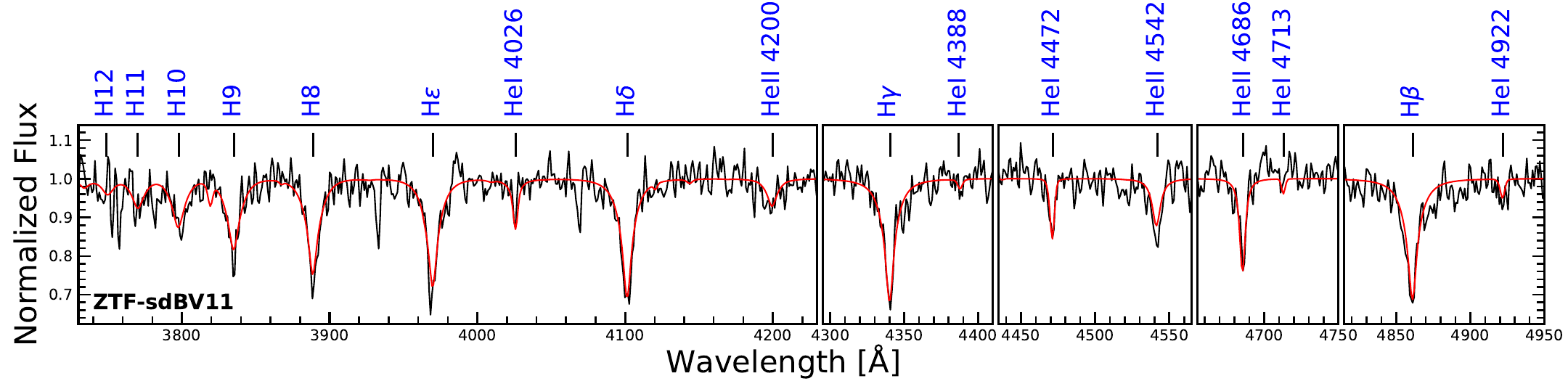}
\caption{Spectroscopic model fitting (red) to the data (black) for ZTF-sdBV11, which we classify as a variable sdO.}.
\label{fig:sdOV_spec}%
\end{figure*}

\begin{figure*}
\centering
\begin{minipage}[t]{0.33\textwidth}
    \centering
    \includegraphics[width=\textwidth]{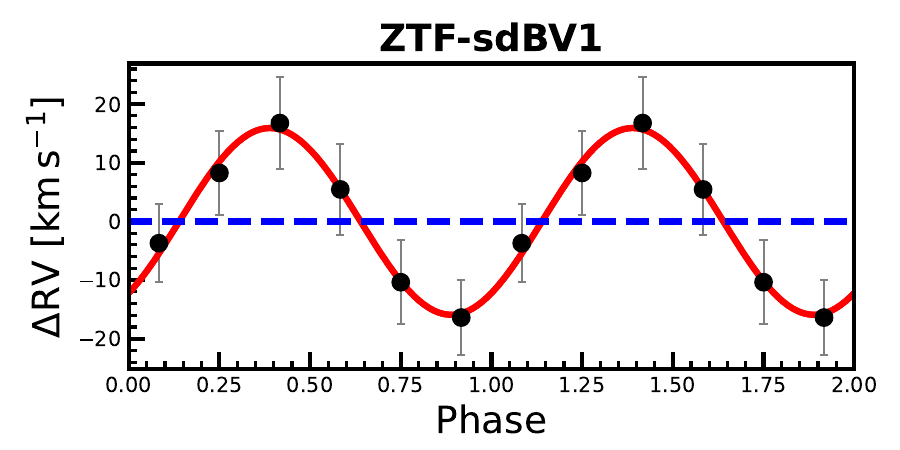}
\end{minipage}%
\hfill
\begin{minipage}[t]{0.33\textwidth}
    \centering
    \includegraphics[width=\textwidth]{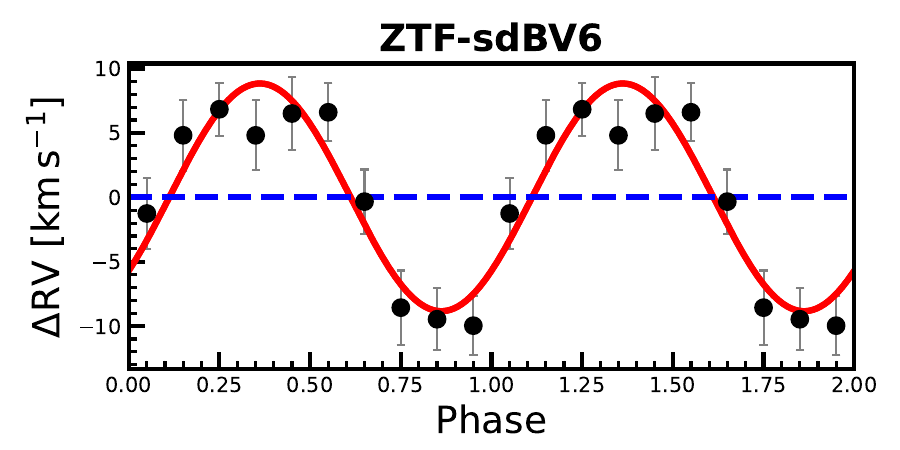}
\end{minipage}
\begin{minipage}[t]{0.33\textwidth}
    \centering
    \includegraphics[width=\textwidth]{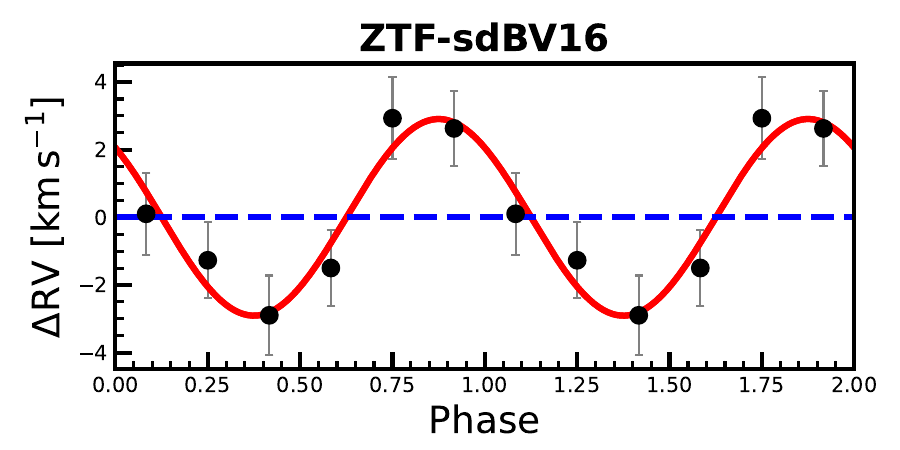}
\end{minipage}
\vspace{1mm}
\caption{Phase-resolved RV results for objects with time-series spectroscopy, which was suitable to find RV variations, but no clear atmospheric variability above the noise level. Measurements (black) are shown, fitted with their single harmonic term (red), around zero RV (blue).}
\label{fig:tsRV}
\end{figure*}

\begin{figure*}[p]
\centering
\begin{minipage}[t]{0.33\textwidth}
    \centering
    \includegraphics[width=\textwidth]{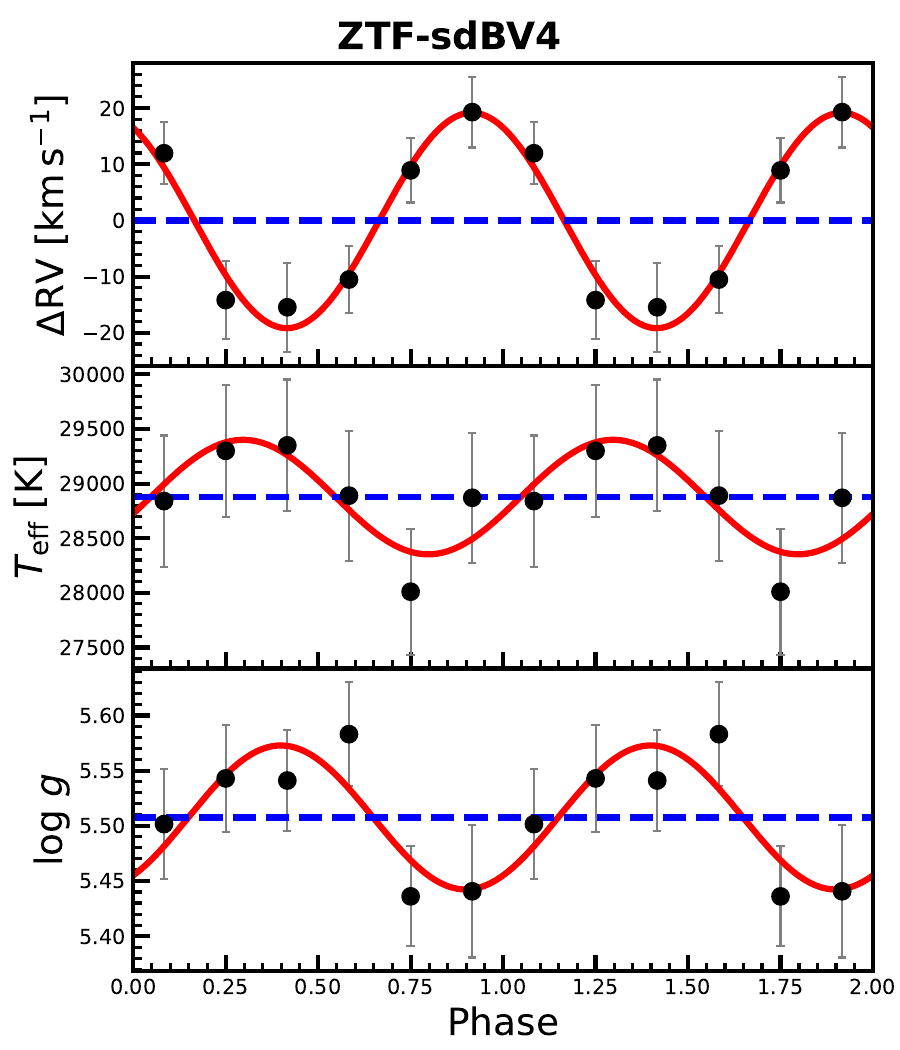}
\end{minipage}%
\hspace{-2mm}
\begin{minipage}[t]{0.33\textwidth}
    \centering
    \includegraphics[width=\textwidth]{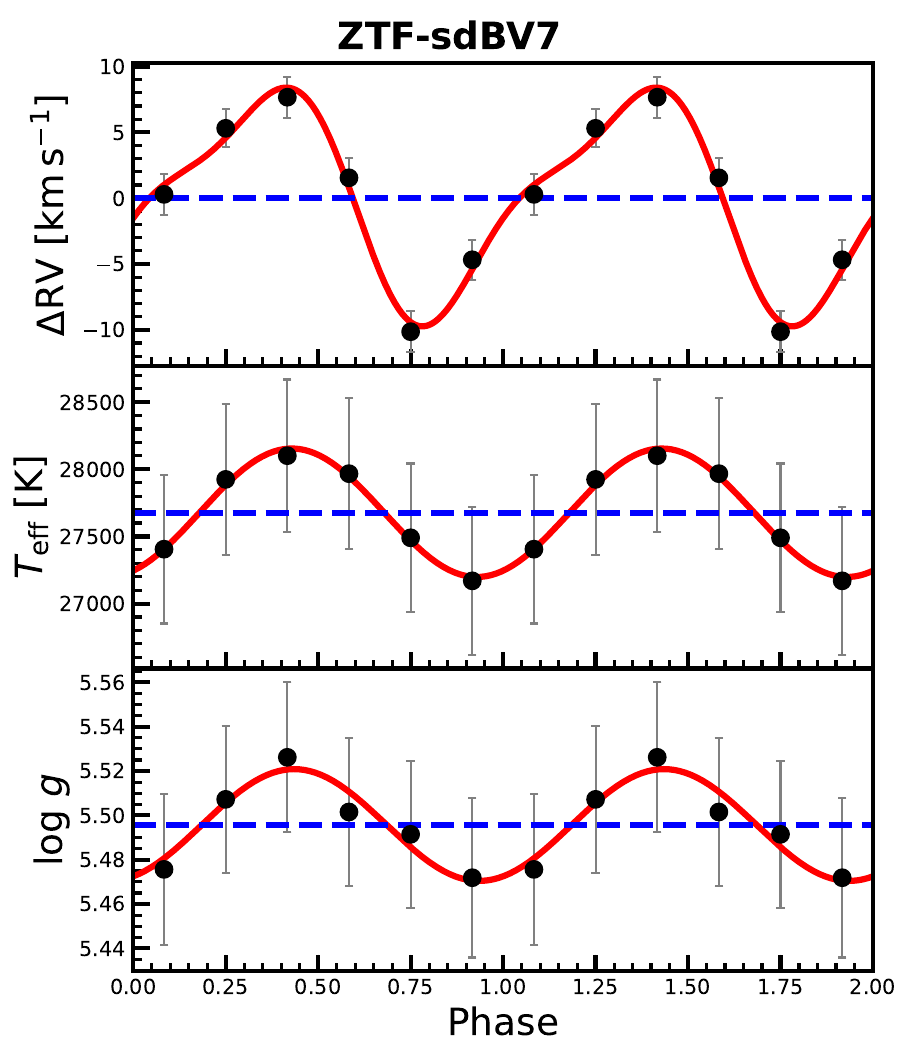}
\end{minipage}
\hspace{-2mm}
\begin{minipage}[t]{0.33\textwidth}
    \centering
    \includegraphics[width=\textwidth]{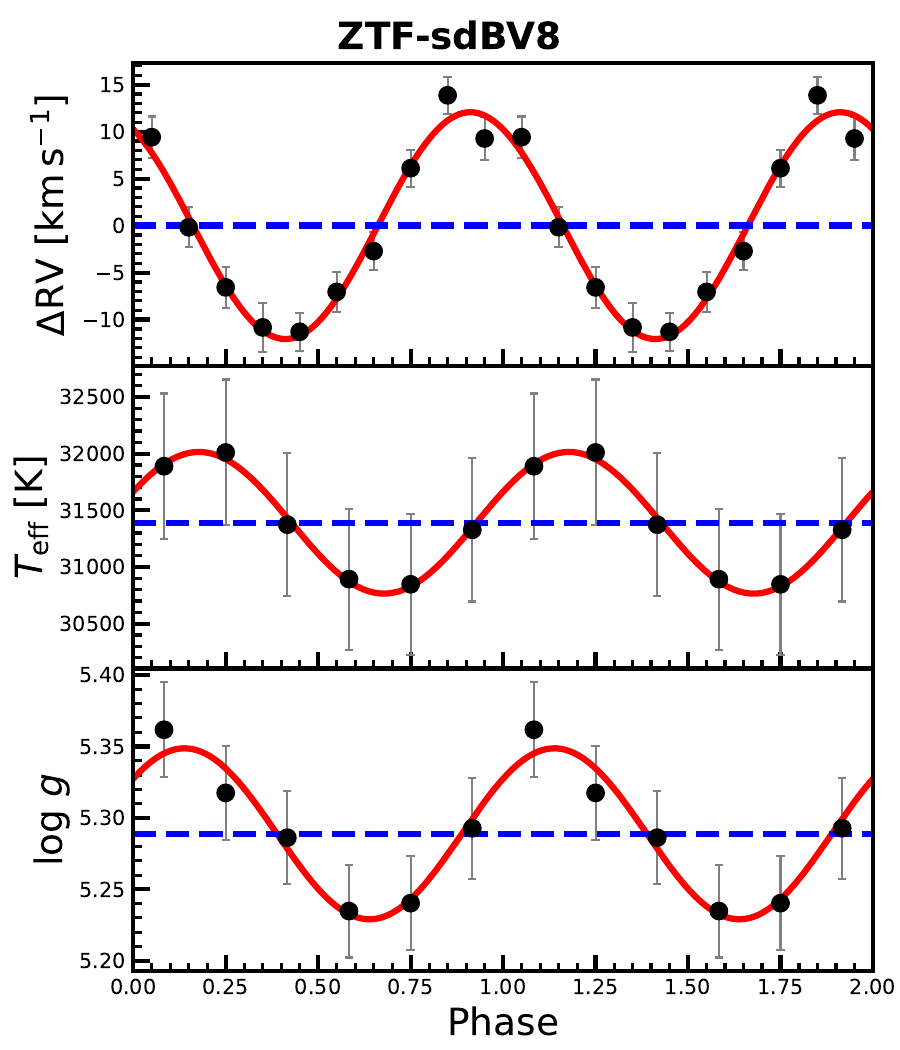}
\end{minipage}
\centering
\includegraphics[width=0.33\textwidth]{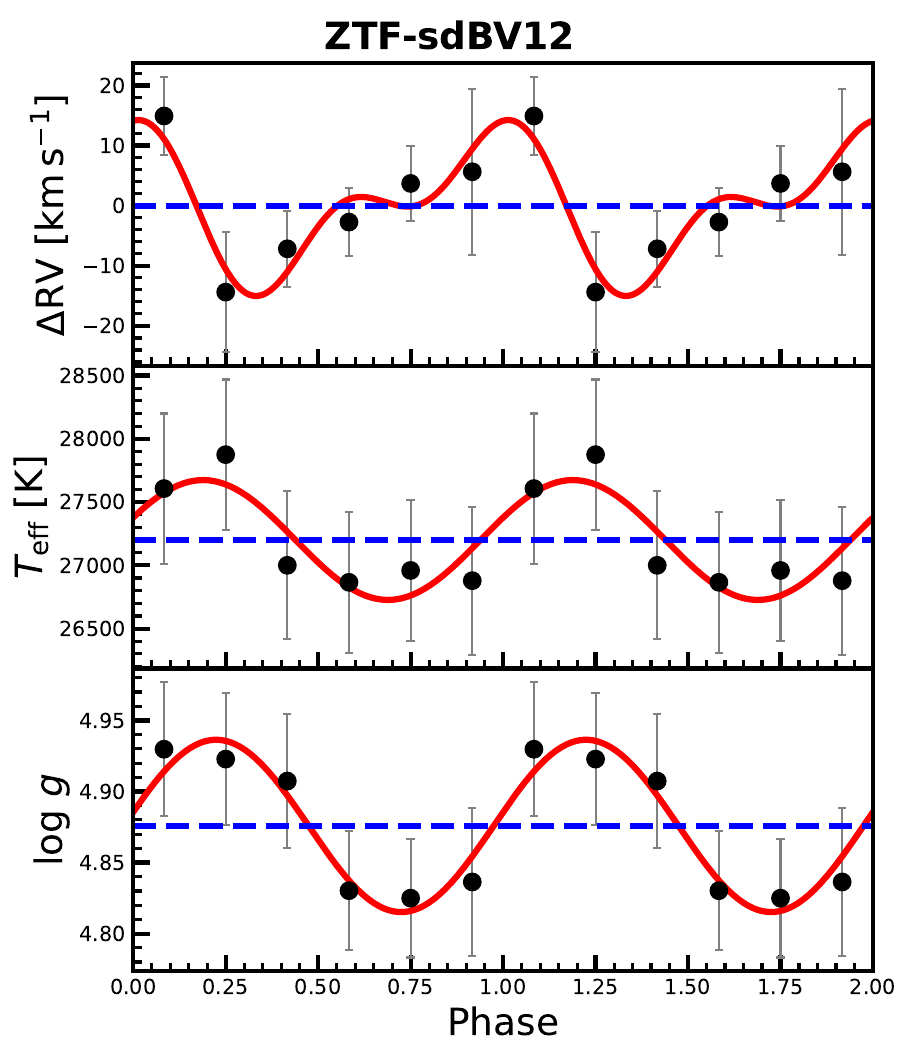} 
\vspace{1mm}
\caption{Phase-resolved RV, {\teff} and {\logg} results for objects with time-series spectroscopy with detected atmospheric variability. Measurements (black) are shown along with their harmonic fitting (red), around the mean value (blue). All data sets have been fitted with with a single harmonic term, except the RV measurements for ZTF-sdBV7 and ZTF-sdBV12, which have been fitted with a second order Fourier term.} 
\label{fig:tsspec}
\end{figure*}

Model fittings to the spectra for each object observed in the current ZTF sample. Fig.~\ref{fig:rmsdBV_spec_grid} shows the fitted spectra for the classified rm-sdBVs. Fig.~\ref{fig:sdBVr_spec} shows the fitted spectra for the two classified sdBV$_r$ stars. Fig.~\ref{fig:BLAP_spec} shows the fitted spectra for the two classified BLAPs. Fig.~\ref{fig:sdOV_spec} shows the spectral fitting for the classified sdOV. For objects which have phase-resolved results, an example fitting for a single bin is shown. Figs.~\ref{fig:tsRV}~and~\ref{fig:tsspec} show the phase-resolved results of the time-series spectral analysis, where phase zero represents the start of the observation baseline for each data set. 

\section{SED fittings} 
\label{sec:apped:sed}
Spectral energy distribution fittings for each object in our sample that has atmospheric {\teff}, {\logg} and {\he} measurements and reasonable parallax errors. The photometric flux measurements are gathered from the following sources. 
\textbf{UV}: Revised catalog of GALEX UV sources (\citealt{SED_UV}; purple). 
\textbf{Optical}: 
SkyMapper Southern Survey (SMSS) DR4 (\citealt{SED_SMSS_DR4}; green),
Pan-STARRS1 survey DR2 (\citealt{SED_PANSTARS1}; dark red), 
SDSS Photometric Catalog Release 12 (\citealt{SED_sdss}; gold),
\textit{Gaia} EDR3 (\citealt{SED_gaia}; light blue) and DR3 XP (\citealt{Gaia_DR3}; black). 
\textbf{Infrared}: Two Micron All-Sky Survey (2MASS) (\citealt{SED_2MASS}; bright red),
Deep Near Infrared Survey of the Southern Sky  (\citealt{SED_DENIS}; dark yellow),
UKIDSS-DR9 LAS, GCS and DXS Surveys (\citealt{SED_ukidss}; maroon),
UHS-DR3 (\citealt{SED_UHS}; light pink)
VISTA Hemisphere Survey (\citealt{SED_vista}; light red),
The band-merged unWISE Catalog (\citealt{SED_unwise}; pink).

\begin{figure*}
\centering
    \includegraphics[width=\textwidth]{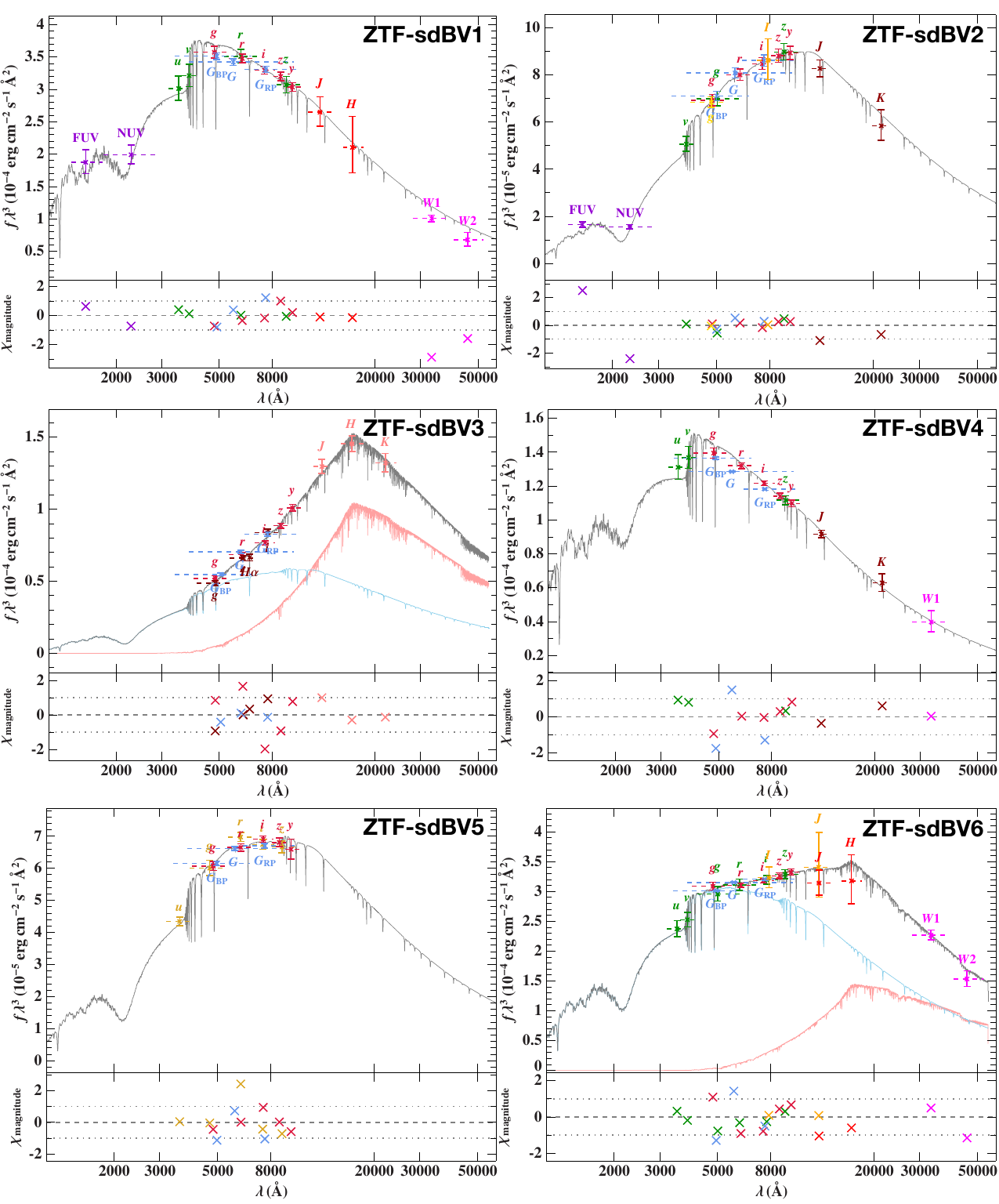}
\end{figure*}
\begin{figure*}
\centering
    \includegraphics[width=\textwidth]{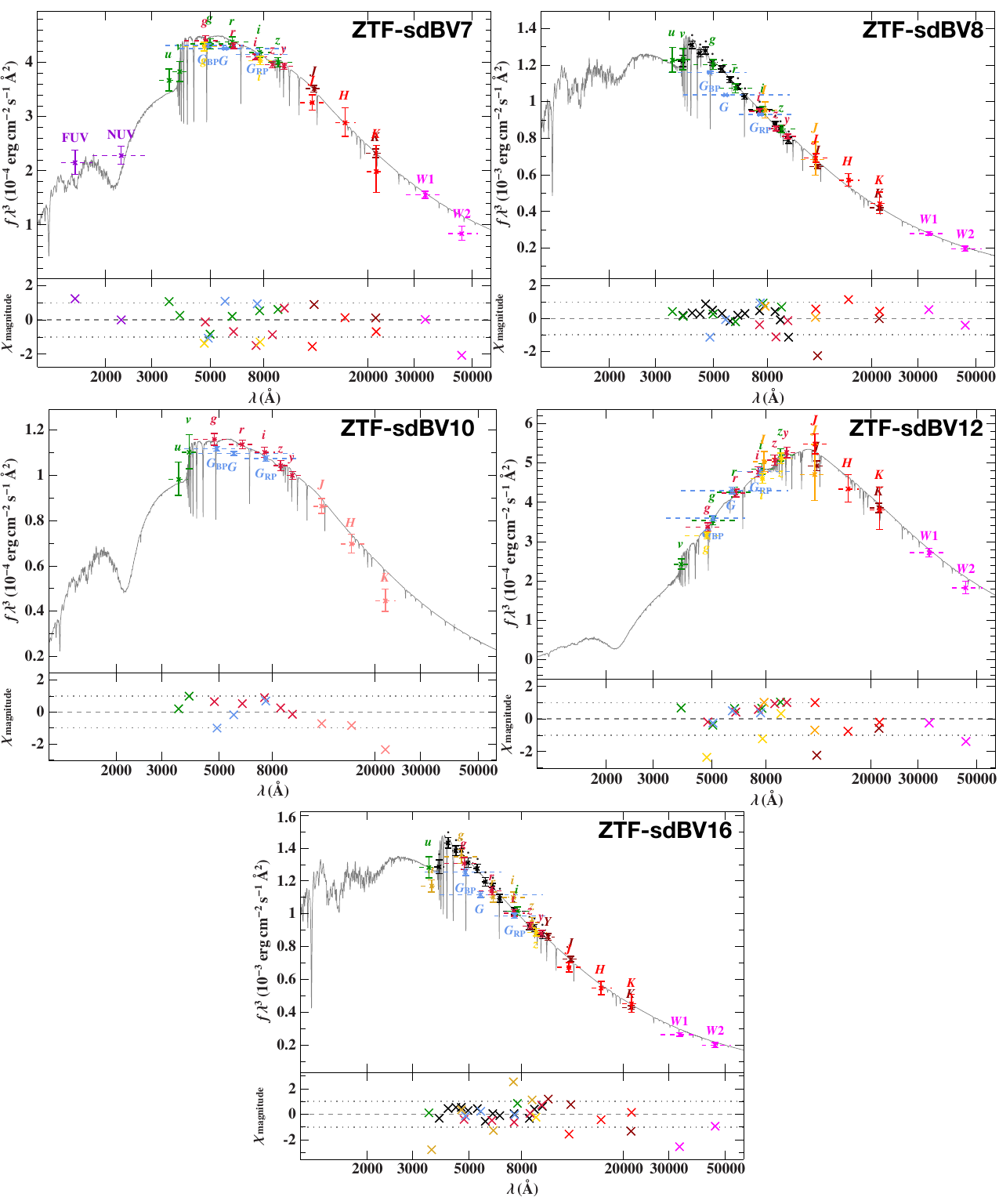}
    \caption{SED fittings used to derive the reported fundamental properties. Photometric fluxes (colored points) are gathered from the sources listed above in Section~\ref{sec:apped:sed}. The best fitting model spectrum (gray) is fitted to the photometric data. For ZTF-sdBV3 and ZTF-sdBV6, a composite model is used consisting of a synthetic sdB spectrum (blue) and model PHEONIX spectrum (red).}
\label{fig:SED_grid2}
\end{figure*} 

\end{appendix}

%
%

\end{document}